\documentclass[12pt,showpacs]{revtex4}
\usepackage[english]{babel}
\usepackage{graphicx}
\usepackage{epsfig}

\usepackage{amsfonts}
\usepackage{amsmath}
\usepackage{latexsym}        
\newcommand{\bea}{\begin{eqnarray}}
\newcommand{\eea}{\end{eqnarray}}
\newcommand{\be}{\begin{equation}}
\newcommand{\ee}{\end{equation}}
\newcommand{\nn}{\nonumber}

\def\ket#1{\hbox{$\vert #1\rangle$}}   
\def\bra#1{\hbox{$\langle #1\vert$}}   

\newcommand{\sign}{{\rm{sign}}}
\newcommand{\slashed}[1]{\rlap{/}{#1}}

\begin{document}


\title{Transverse momentum dependent parton distributions in a light-cone quark model}
\normalsize
\author{B. Pasquini}
\author{S. Cazzaniga}
\author{S. Boffi}
\affiliation{%
Dipartimento di Fisica Nucleare e Teorica, Universit\`{a} degli Studi di Pavia, and\\
Istituto Nazionale di Fisica Nucleare, Sezione di Pavia, I-27100 Pavia, Italy}


\begin{abstract}
The leading twist transverse momentum dependent parton distributions (TMDs) are studied in a light-cone description of the nucleon where the Fock expansion is truncated to consider only valence quarks. General analytic expressions are derived in terms of the six amplitudes needed to describe the three-quark sector of the nucleon light-cone wave function. Numerical calculations for the T-even TMDs are presented in a light-cone constituent quark model, and the role of the so-called pretzelosity is investigated to produce a nonspherical shape of the nucleon.
\end{abstract}

\date{\today}

\maketitle


\section{Introduction}
\label{sect:intro}

In recent years much work has been devoted to study semi-inclusive deep inelastic scattering (SIDIS) and Drell-Yan (DY) dilepton production as powerful tools to understand the nucleon structure. According to the factorization theorem that separates the coherent long distance interactions between projectile and target from the incoherent short distance interactions~\cite{Collins:1981uk,Ji:2004wu,Collins:2004nx},  the physical observables of such processes can be expressed as convolution of hard partonic scattering cross sections, parton distribution functions (PDFs) and parton fragmentation functions (FFs)~\cite{Collins:2003fm,Collins:2007ph,Bom08}. With respect to the usual inclusive deep inelastic scattering (DIS) where PDFs only depend on the longitudinal momentum fraction carried by the parton, now PDFs, as well as FFs, also depend on the transverse momentum. At leading twist there are eight transverse momentum dependent PDFs (TMDs)~\cite{Mulders2,BoerMulders,Goeke,Bacchetta}, three of them surviving when integrated over the transverse momentum and giving rise to the familiar parton density, helicity and transversity distributions. 

TMDs contain rich and direct three-dimensional information about the internal dynamics of the nucleon. In particular, they can help in understanding how the nucleon spin originates from the quark spins and the orbital angular momentum of quarks and gluons. However, being typical nonperturbative quantities, TMDs are not directly calculable in quantum chromodynamics (QCD), and their modeling requires assumptions about the nucleon wave function.

When dealing with high-energy scattering where hadrons travel near the speed of light, the most natural tool to describe the nucleon is the light-cone Fock state expansion of its wave function. The hadronic state is decomposed in terms of  $N$-parton Fock states with coefficients representing the momentum light-cone wavefunction (LCWF) of the $N$ partons~\cite{LB80,BPP98}. In principle there is an infinite number of LCWFs in such an expansion. However, since the constituent quark models work so well phenomenologically, there must exist a light-cone description of the nucleon in which only the Fock components with a few partons are necessary.

The quark distribution amplitudes defined in terms of hadron-to-vacuum transition matrix elements of non-local gauge-invariant light-cone operators and describing the three-quark component of the nucleon have been studied extensively in the literature~\cite{LB80,Chernyak:1984ej,King:1987wi,Chernyak:1989nv,Braun:1999te,Braun:2000kw,Stefanis}. In turn, the light-cone Fock expansion of a hadron state is completely defined by the matrix elements of a special  class equal light-cone time quak-gluon operators between the QCD vacuum $\vert 0\rangle$ and the hadron~\cite{BurkJiY02}. The authors of Refs.~\cite{Ji:2002xn,Ji:2004,Ji:2003} considered the wave-function amplitudes keeping full transverse-momentum dependence of partons and proposed a systematic way to enumerate independent amplitudes of a LCWF given a particular parton combination. If one truncates the light-cone expansion of the proton state to the minimum Fock sector with just three valence quarks, one can write down the  matrix elements of a class of three-quark light-cone operators which serve to define a complete set of light-cone amplitudes within the truncation.  These matrix elements can be simplified using color, flavor, spin and discrete symmetries~\cite{Ji:2002xn,Ji:2004,Ji:2003}, and at the end one finds that six amplitudes are needed to describe the three-quark sector of the nucleon LCWF. Depending on the imposed gauge fixing conditions such amplitudes are real or complex. In the latter case, the amplitudes contain final state interaction effects.

With such amplitudes one can calculate nucleon observables. One could choose a phenomenological approach parametrizing and fitting them to data. Here, the wave-function amplitudes are modeled in a light-cone constituent quark model (CQM) which has been successfully applied in the calculation of the electroweak properties of the nucleon~\cite{PBff}, generalized parton distributions~\cite{BPT1,BPT2,PPB,PBcloud} and spin densities~\cite{PBspin}. This representation is well suited to disentangle the contribution from the different orbital angular momentum components of the nucleon wave function, and therefore to study the spin-spin and spin-orbit correlations encoded in the different TMDs.

The paper is organized as follows. After a brief review of the definition of TMDs in Section~\ref{sect:tmd} and the resulting light-cone amplitudes when considering only three active quarks in Section~\ref{sect:ampli}, the proton wave function is constructed in Section~\ref{sect:lcwf} where the representation for the nucleon amplitudes in terms of the light-cone CQM is derived. Time-even TMDs are then calculated in Section~\ref{sect:tmdlc} confining explicit expressions for the light-cone amplitude overlap representation of both time-even and time-odd TMDs in Appendix A. Numerical results are presented in Section~\ref{sect:result}, and concluding remarks are given in Section~\ref{sect:conclude}.


\section{Transverse momentum dependent parton distributions}
\label{sect:tmd}

In this section we review the formalism for the definition of TMDs, following the conventions of Refs.~\cite{Mulders2,BoerMulders,Goeke,Bacchetta}. The quark-quark distribution correlation function appearing in SIDIS is defined as 
\begin{eqnarray}
\Phi_{ij}(x,\boldsymbol{k}_{\perp},S)
=\int\frac{{\rm{d}}\xi^{-}{\rm{d}}^{2}\boldsymbol{ \xi}_{\perp}}{(2\pi)^{3}}
e^{i(k^{+}\xi_{-}-\boldsymbol{ k}_{\perp}\cdot\boldsymbol{ \xi}_{\perp})}
\left.\langle
P,S|\overline{\psi}_{j}(0) 
{\cal U}^{n_-}_{(0,+\infty)}
{\cal U}^{n_-}_{(+\infty,\xi)}
\psi_{i}(\xi)|P,S\rangle\right|_{\xi^{+}=0} ,
\label{eq:31}
\end{eqnarray}
where $k^+=xP^+$, and here and in the following we omit flavor indices. ${\cal U}$ is the Wilson link connecting the two quark fields and ensuring color gauge invariance of the correlator~\cite{Bom04}. The target state is characterized by its four-momentum $P$ and the covariant spin vector $S=(S^+,S^-,\boldsymbol{S}_\perp)$ ($P^2=M^2, S^2=-1, P\cdot S=0$), where
\begin{eqnarray}
S^+= \Lambda\,\frac{P^+}{M},\quad S^-=-\Lambda\,\frac{P^-}{M}.
\label{eq:32}
\end{eqnarray}

The TMDs  enter the general decomposition of the correlator 
$\Phi_{ij}(x,\boldsymbol{k}_{\perp},S)$ which, at the twist-two level, reads
\begin{eqnarray}
\Phi(x,\boldsymbol{k}_{\perp},S)&=&
\frac{1}{2}\left\{
f_{1}\slashed{n}_{+}
-f_{1T}^{\perp}\frac{\epsilon^{ij}_{T}\boldsymbol{k}_{\perp}^iS_{\perp}^j}{M}\slashed{n}_{+}
+\Lambda g_{1L}\gamma_{5}\slashed{n}_{+}\right.\nonumber\\
&&\qquad+\left.\frac{(\boldsymbol{k}_{\perp}\cdot
\boldsymbol{S}_{\perp})}{M}g_{1T}\gamma_{5}\slashed{n}_{+}
+h_{1T}\frac{[\slashed{S}_{\perp},\slashed{n}_{+}]}{2}\gamma_{5}\right.\nonumber\\
&&\qquad +\Lambda h_{1L}^{\bot}
 \frac{[\slashed{{k}}_{\perp},\slashed{n}_{+}]}{2M}\gamma_{5}
+\left.\frac{({\boldsymbol{k}}_{\perp}\cdot
{\boldsymbol{S}}_{\perp})}{M}h_{1T}^{\bot}\frac{[\slashed{{k}}_{\perp},
\slashed{n}_{+}]}{2M}\gamma_{5}\right.\nonumber\\
&&\qquad+\left.ih_{1}^{\perp}\frac{[\slashed{{k}}_{\perp},\slashed{n}_{+}]}{2M}\right\},\label{eq:33}
\end{eqnarray}
where $n_+$ and $n_-$ are two lightlike vectors satisfying $n_+\cdot n_-=1$, 
and $ \epsilon_T^{ij}=\epsilon^{-+ij}$, and the transverse four-vectors are defined as $v_\perp=(0,0,\boldsymbol{v}_\perp)$. The nomenclature of the distribution functions follows closely that
of Ref.~\cite{Mulders2}, sometimes referred to as ``Amsterdam notation'': $f$ refers to unpolarized target, $g$ and $h$ to longitudinally and transversely polarized target, respectively; a subscript $1$ is given to the twist-two functions, subscripts $L$ or $T$ refer to the connection with the hadron spin being longitudinal or transverse and a symbol $\perp$ signals the explicit presence of transverse momenta with an uncontracted index. Among the eight distributions of Eq.~(\ref{eq:33}), the Boer-Mulders TMD $h_1^\perp$~\cite{BoerMulders} and the Sivers function $f_{1T}^\perp$~\cite{Sivers} are T-odd, i.e. they change sign under ``naive time reversal'', which is defined as usual time reversal, but without interchange of initial and final states. All the TMDs in Eq. (\ref{eq:33}) depend on $x$ and $\boldsymbol{k}^{2}_{\perp}$. These functions can be individually isolated by performing traces of the correlator with suitable Dirac matrices. Using the abbreviation $\Phi^{[\Gamma]}\equiv{\rm Tr}(\Phi\Gamma)/2$, and restricting ourselves to the T-even TMDs, we have 
\begin{center}
\begin{eqnarray}
\label{eq:34}
\Phi^{[\gamma^{+}]}(x, \boldsymbol{k}_{\perp})
&=&f_{1}, \\ 
\label{eq:35}
\Phi^{[\gamma^{+}\gamma_{5}]}(x, \boldsymbol{k}_{\perp})
&=&\Lambda g_{1L}
+ \frac{(\boldsymbol{k}_{\perp}\cdot \boldsymbol{S}_{\perp})}{M} \,g_{1T}, \\
\label{eq:36}
\Phi^{[i\sigma^{j+}\gamma_{5}]}(x, \boldsymbol{k}_{\perp})
&=&{S}_{\perp}^{j}h_{1T}
+\frac{k_{\perp}^j}{M}\left[\Lambda h_{1L}^{\perp}
+\frac{(\boldsymbol{k}_{\perp}\cdot \boldsymbol{S}_{\perp})}{M}\,h_{1T}^{\perp}\right], \\
\label{eq:37}
&=&{S}_{\perp}^{j}h_{1}
+\Lambda\frac{ k_{\perp}^j}{M}\,h_{1L}^{\perp}
+S_{\perp}^i \frac
{2k^{i}_{\perp}k^{j}_{\perp} -\boldsymbol{k}^{2}_{\perp} \delta^{ij}} {2M^{2}} \, h_{1T}^{\perp} ,
\end{eqnarray}
\end{center}
where $j=x,y$ is a transverse index, and
\begin{equation}
h_{1}=h_{1T}+\frac{\boldsymbol{k}^{2}_{\perp}}{2M^2}\,h_{1T}^{\perp}.
\label{eq:38}
\end{equation} 

The correlation function $\Phi^{[\gamma^{+}]}(x,\boldsymbol{k}_{\perp})$ is just the unpolarized quark distribution, which integrated over $\boldsymbol{k}_{\perp}$ gives the familiar light-cone momentum distribution $f_{1}(x)$. All the other TMDs characterize the strength of different spin-spin and spin-orbit correlations. The precise form of this correlation is given by the prefactors of the TMDs in 
Eqs.~(\ref{eq:35})-(\ref{eq:37}). In particular, the TMDs $g_{1L}$ and $h_1$ describe the strength of a correlation between a longitudinal/transverse target polarization and a longitudinal (circular)/transverse (linear) parton polarization. After integration over $\boldsymbol{k}_\perp$, they reduce to the helicity and transversity distributions, respectively. By definition the spin-orbit correlations described by $g_{1T}$, $h_{1L}^\perp$ and $h_{1T}^\perp$ involve the transverse parton momentum and the polarization of both the parton and the target, and vanish upon integration over $\boldsymbol{k}_\perp$.

If one calculates these distributions in the light-cone gauge $A^+=0$ using the advance boundary condition for the transverse component of the gauge field, the gauge links in the quark-quark correlator can be ignored~\cite{JiYuan:02,JiYuanBel:03,Mulders}. However, in this case the wave function amplitudes are not real, and apart from the structural information on the hadron it has also an imaginary phase mimicking the final state interactions. 

By using for the quark fields the canonical expansion in terms of Fock operators, one can write the T-even TMDs as 
\begin{eqnarray}
\label{eq:39}
&&f_{1}^q(x, \boldsymbol{ k}^2_\perp)
=
\langle P \Lambda|\sum_{\lambda} q^\dagger_{\lambda}(\tilde k) q_{\lambda}(\tilde k)|P \Lambda\rangle,\\
&&\nn
\\
\label{eq:40}
&&\Lambda\, g_{1L}^q(x, \boldsymbol{ k}^2_\perp)
=
\langle P\Lambda|\sum_{\lambda}(-1)^{(\frac{1}{2}-\lambda)}
q^\dagger_{\lambda}(\tilde k)q_{\lambda}(\tilde k)|P\Lambda\rangle,
\\
&&\nn
\\
\label{eq:41}
&&\frac{(\boldsymbol{k}_{\perp} \cdot \boldsymbol{S}_{\perp})}{M}\,g_{1T}^q(x, \boldsymbol{ k}^2_\perp)
=
\langle PS_{\bot}|\sum_{\lambda}(-1)^{(\frac{1}{2}-\lambda)}
q^\dagger_{\lambda}(\tilde k)q_{\lambda}(\tilde k)|PS_{\bot}\rangle ,
\\
&&\nn
\\
\label{eq:42}
&&h_1^q(x, \boldsymbol{ k}^2_\perp) = \frac{1}{2}
\sum_\lambda\left[
\langle PS^x|
q^\dagger_{-\lambda}(\tilde k)q_{\lambda}(\tilde k)|PS^x\rangle
+i
\langle PS^y|[\sign(\lambda)]
q^\dagger_{-\lambda}(\tilde k)q_{\lambda}(\tilde k)|PS^y\rangle\right],
\\
&&\nn
\\
\label{eq:43}
&&\Lambda\frac{k_{\perp}^{j}}{M}\,h_{1L}^{\bot\,q}(x, \boldsymbol{k}^2_\perp)
=
-(i)^{j+1}
\langle P\Lambda|\sum_{\lambda}
[\sign(\lambda)]^{j+1}
q^\dagger_{-\lambda}(\tilde k)q_{\lambda}(\tilde k)|P\Lambda\rangle,
\\
&&\nn
\\
\label{eq:44}
&&\frac{(k^2_x-k^2_y)}{M^2}\,h_{1T}^{\perp\,q}(x, \boldsymbol{ k}^2_\perp)
=
\sum_\lambda\left[
\langle PS^x|
q^\dagger_{-\lambda}(\tilde k)q_{\lambda}(\tilde k)|PS^x\rangle \right.\nn\\
&&\hspace{5cm} 
\left. -i
\langle PS^y|[\sign(\lambda)]
q^\dagger_{-\lambda}(\tilde k)q_{\lambda}(\tilde k)|PS^y\rangle\right],
\end{eqnarray}
where  $x>0$, and $q_\lambda(\tilde k)$ ($q^\dagger_\lambda(\tilde k)$) is the annihilation (creation) operator of a quark of flavor $q$ with helicity $\lambda$ and momentum  $\tilde k=(k^+,\boldsymbol{k}_\perp)$.


\section{Three-quark light-cone amplitudes}
\label{sect:ampli}

In this section we reproduce the main results of Ref.~\cite{Ji:2002xn} for the classification of the three-quark LCWF of the nucleon with helicity $\Lambda$ in terms of the total parton light-cone helicity $\lambda$, or, equivalently, in terms of the angular momentum projection $l_z=\Lambda-\lambda$ which follows from angular momentum conservation. For a proton with helicity $\Lambda=1/2$, the complete three-quark light-cone Fock expansion has the following structure
\begin{eqnarray}
|P\uparrow\rangle=|P\uparrow\rangle^{l_z=0}
+|P\uparrow\rangle^{l_z=1}
+|P\uparrow\rangle^{l_z=-1}
+|P\uparrow\rangle^{l_z=2}.
\label{eq:1}
\end{eqnarray}
The different angular-momentum components of the state in Eq.~(\ref{eq:1}) are given by
\begin{eqnarray}
  |P\uparrow\rangle^{l_z=0} &=& \int d[1]d[2]d[3]\left( \psi^{(1)}(1,2,3)
         + i\epsilon_T^{\alpha\beta}k_{1\alpha}k_{2\beta}  \psi^{(2)}(1,2,3)\right) \nonumber \\
         &&  \times  \frac{\epsilon^{ijk}}{\sqrt{6}} u^{\dagger}_{i\uparrow}(1)
             \left(u^{\dagger}_{j\downarrow}(2)d^{\dagger}_{k\uparrow}(3)
            -d^{\dagger}_{j\downarrow}(2)u^{\dagger}_{k\uparrow}(3)\right)
         |0\rangle \ ,
  \label{eq:3} \\
&&\nonumber\\
  |P\uparrow\rangle^{l_z=1} &=& \int d[1]d[2]d[3]\left(k_{1\perp}^+
           \psi^{(3)}(1,2,3)
         + k_{2\perp}^+ \psi^{(4)}(1,2,3)\right) \nonumber \\
         &&  \times  \frac{\epsilon^{ijk}}{\sqrt{6}} \left( u^{\dagger}_{i\uparrow}(1)
            u^{\dagger}_{j\downarrow}(2)d^{\dagger}_{k\downarrow}(3)
            -d^{\dagger}_{i\uparrow}(1)u^{\dagger}_{j\downarrow}(2)
             u^{\dagger}_{k\downarrow}(3)\right)
         |0\rangle \ ,
   \label{eq:4}\\
&&\nonumber\\
  |P\uparrow\rangle^{l_z=-1} &=& \int d[1]d[2]d[3](-)k_{2\perp}^-
          \psi^{(5)}(1,2,3) \nonumber \\
         &&  \times  \frac{\epsilon^{ijk}}{\sqrt{6}} u^{\dagger}_{i\uparrow}(1)
             \left(
     u^{\dagger}_{j\uparrow}(2)d^{\dagger}_{k\uparrow}(3)
    -d^{\dagger}_{j\uparrow}(2)u^{\dagger}_{k\uparrow}(3)
            \right)
         |0\rangle \ ,
    \label{eq:5}\\
&&\nonumber\\
  |P\uparrow\rangle^{l_z=2} &=& \int d[1]d[2]d[3]~k_{1\perp}^+
k_{3\perp}^+
         \psi^{(6)}(1,2,3)  \nonumber \\
         &&  \times  \frac{\epsilon^{ijk}}{\sqrt{6}} u^{\dagger}_{i\downarrow}(1)
             \left(d^{\dagger}_{j\downarrow}(2)u^{\dagger}_{k\downarrow}(3)-u^{\dagger}_{j\downarrow}(2)d^
{\dagger}_{k\downarrow}(3)
            \right)
         |0\rangle \ ,
   \label{eq:6}
\end{eqnarray}
where $\alpha,\beta=1,2$ are transverse indexes and $k^\pm_{i\perp}=k^x_i\pm k^y_i$.
In Eqs.~(\ref{eq:3})-(\ref{eq:6}), the integration measures are defined as
\begin{equation}
\label{eq:7}
d[1]d[2]d[3]=
\frac{dx_1dx_2dx_3}{\sqrt{x_1x_2x_3}}\delta\left(1-\sum_{i=1}^3 x_i\right)
\frac{d^2 \boldsymbol{k}_{1\perp}d^2\boldsymbol{k}_{2\perp}
d^2\boldsymbol{k}_{3\perp}}{[2(2\pi^3)]^2}
\delta\left(\sum_{i=1}^3 \boldsymbol{k}_{\perp\,i}\right),
\end{equation}
where $x_i$ are the fraction of the longitudinal nucleon momentum carried by the quarks, and
$\boldsymbol{k}_{i\perp}$ are their transverse momenta. Furthermore, $u^\dagger_{i\lambda}$ ($u_{i\lambda}$) and $d^\dagger_{i\lambda}$ ($d_{i\lambda}$) are creation (annihilation) operators of up and down quarks with helicity $\lambda$ and color $i$, respectively, and $\psi^{(j)}$ are functions
of quark momenta with argument $i$ representing $x_i$ and $\boldsymbol{k}_{i\perp}$, and the dependence on the transverse momenta  is of the form $\boldsymbol{k}_{i\perp}\cdot \boldsymbol{k}_{j\perp}$ only.

The proton state with negative helicity is given in terms of the same wave function amplitudes $\psi^{(j)}$, except that the quark helicities are flipped, $k^x\pm ik^y$ become  $k^x\mp ik^y$, and some signs are added 
\begin{eqnarray}
  |P\downarrow\rangle^{l_z=0} &=& \int d[1]d[2]d[3]
\left( -\psi^{(1)}(1,2,3)+ i\epsilon_T^{\alpha\beta}k_{1\alpha}k_{2\beta}  \psi^{(2)}(1,2,3)\right) \nonumber \\
         &&  \times  \frac{\epsilon^{ijk}}{\sqrt{6}} u^{\dagger}_{i\downarrow}(1)
             \left(u^{\dagger}_{j\uparrow}(2)d^{\dagger}_{k\downarrow}(3)
            -d^{\dagger}_{j\uparrow}(2)u^{\dagger}_{k\downarrow}(3)\right)
         |0\rangle \ ,
   \label{eq:8} \\
&&\nonumber\\
  |P\downarrow\rangle^{l_z=-1} &=& \int d[1]d[2]d[3]\left(k_{1\perp}^-
           \psi^{(3)}(1,2,3)
         + k_{2\perp}^- \psi^{(4)}(1,2,3)\right) \nonumber \\
         &&  \times  \frac{\epsilon^{ijk}}{\sqrt{6}} 
	 \left( u^{\dagger}_{i\downarrow}(1)
         u^{\dagger}_{j\uparrow}(2)d^{\dagger}_{k\uparrow}(3)
         -d^{\dagger}_{i\downarrow}(1)u^{\dagger}_{j\uparrow}(2)
         u^{\dagger}_{k\uparrow}(3)\right)
         |0\rangle \ ,
    \label{eq:9}\\
&&\nonumber\\
  |P\downarrow\rangle^{l_z=1} &=& \int d[1]d[2]d[3]~(-)k_{2\perp}^+
          \psi^{(5)}(1,2,3) \nonumber \\
         &&  \times  \frac{\epsilon^{ijk}}{\sqrt{6}} u^{\dagger}_{i\downarrow}(1)
             \left(
     u^{\dagger}_{j\downarrow}(2)d^{\dagger}_{k\downarrow}(3)
    -d^{\dagger}_{j\downarrow}(2)u^{\dagger}_{k\downarrow}(3)
            \right)
         |0\rangle \ ,
    \label{eq:10}\\
&&\nonumber\\
	 |P\downarrow\rangle^{l_z=-2} &=& 
	 \int d[1]d[2]d[3](-)~k_{1\perp}^-k_{3\perp}^-
         \psi^{(6)}(1,2,3)  \nonumber \\
         &&  \times  \frac{\epsilon^{ijk}}{\sqrt{6}} 
	 u^{\dagger}_{i\uparrow}(1)
         \left(d^{\dagger}_{j\uparrow}(2)u^{\dagger}_{k\uparrow}(3)
	 -u^{\dagger}_{j\uparrow}(2)d^
	 {\dagger}_{k\uparrow}(3)
         \right)
         |0\rangle \ .
   \label{eq:11}
\end{eqnarray}

As the previous results are derived within light-front quantization, we implicitly assumed to work in the light-cone gauge $A^+=0$. However, this last condition does not fix the gauge completely, and additional boundary conditions must be specified. Depending on whether the additional gauge condition satifies time reversal or not, the wave function amplitudes are real or complex~\cite{Ji:2002xn} (see also~\cite{CS08}).


\section{Nucleon wave function in a light-cone constituent quark model}
\label{sect:lcwf}

In this section we derive the light-cone amplitudes $\psi^{(i)}$ in a light-cone CQM. Working in the so-called ``uds'' basis~\cite{Franklin:68,Capstick:86} the proton state is given in terms of a completely symmetrized wave function of the form
\begin{equation}
|P\uparrow\rangle=|P\uparrow\rangle_{uud}+|P\uparrow\rangle_{udu}+
|P\uparrow\rangle_{duu} \,.
\label{eq:2}
\end{equation}
In this symmetrization, the state $|P\uparrow\rangle_{udu}$ is obtained from
$|P\uparrow\rangle_{uud}$ by interchanging the second and third spin and space coordinates as well as the indicated quark type, with a similar interchange of the first and third coordinates for 
$|P\uparrow\rangle_{duu}$. 

Following the derivation outlined in Ref.~\cite{BPT1}, we find that the $uud$ component of the light-cone state of the proton can be written as
\be\label{eq:12}
\ket{P,\Lambda}_{uud} = \sum_{\lambda_i,c_i}
\int d[1]d[2]d[3]
\Psi^{\Lambda,[f]}_{uud}(\{x_i,\boldsymbol{ k}_{\perp i};\lambda_i\})
\frac{\epsilon^{ijk}}{\sqrt{6}} 
u^{\dagger}_{i\lambda_1}(1)
u^{\dagger}_{j\lambda_2}(2)
d^{\dagger}_{k\lambda_3}(3)
|0\rangle,
\ee
where assuming SU(6) spin-flavor symmetry the LCWF
$\Psi^{\Lambda,[f]}_{uud}(\{x_i,\boldsymbol{ k}_{\perp i};\lambda_i\})$ 
is given by
\begin{eqnarray}
\label{eq:13}
   \Psi^{\Lambda,[f]}_{uud}(\{x_i,\boldsymbol{ k}_{\perp i}; \lambda_i\}) 
&=& 
\tilde \psi(\{x_i,\boldsymbol{ k}_{\perp i}\})
\frac{1}{\sqrt{3}}\tilde\Phi_{\Lambda}(\lambda_1,\lambda_2,\lambda_3).
\end{eqnarray}

In Eq.~(\ref{eq:13}) the momentum dependent wave function is defined as
\begin{eqnarray}\label{eq:14}
\tilde \psi(\{x_i,\boldsymbol{ k}_{\perp i}\})=
2(2\pi)^3\bigg[\frac{1}{M_0}\frac{\omega_1\omega_2\omega_3}{x_1x_2x_3}\bigg]^{1/2}\psi(\{x_i,\boldsymbol{ k}_{\perp i}\}),
\end{eqnarray}
with $\psi(\{x_i,\boldsymbol{ k}_{\perp i}\})$ symmetric under exchange of the momenta of any quark pairs, $\omega_i$ the free-quark energy, and $M_0=\sum_{i=1}^3 \omega_i$ the mass of the non-interacting three-quark system. The spin dependent part in Eq.~(\ref{eq:13}) is given by
\begin{eqnarray}
\tilde\Phi_{\Lambda}(\lambda_1,\lambda_2,\lambda_3)
&=&\sum_{\mu_1\mu_2\mu_3}
\langle 1/2,\mu_1; 1/2, \mu_2|1, \mu_1+\mu_2 \rangle
\langle 1, \mu_1+\mu_2;1/2, \mu_3| 1/2, \Lambda\rangle\nn\\
&&\times 
D_{\mu_1\lambda_1}^{1/2*}(R_{cf}(x_1,\boldsymbol{ k}_{\perp 1}))
D_{\mu_2\lambda_2}^{1/2*}(R_{cf}(x_2,\boldsymbol{ k}_{\perp 2}))
D_{\mu_3\lambda_3}^{1/2*}(R_{cf}(x_3,\boldsymbol{ k}_{\perp 3})),
\label{eq:15}
\end{eqnarray}
where $D_{\lambda\mu}^{1/2}(R_{cf}(x,\boldsymbol{ k}_\perp))$ is a matrix element of the Melosh rotation $R_{cf}$~\cite{Melosh:74},
\begin{eqnarray}
D_{\lambda\mu}^{1/2}(R_{cf}(x,\boldsymbol{ k}_\perp)) &=&
\langle\lambda|R_{cf}(x,\boldsymbol{k}_\perp)|\mu\rangle\nonumber\\
&=&
\langle\lambda|\frac{m + xM_0 -
i\boldsymbol{\sigma}\cdot(\hat{\boldsymbol{z}}\times\boldsymbol{k}_\perp)}{\sqrt{(m
+ xM_0)^2 + \boldsymbol{k}_\perp^2}}|\mu\rangle.
\label{eq:16}
\end{eqnarray}
The Melosh rotation corresponds to the unitary transformation which converts the Pauli spinors of the quark in the nucleon rest-frame to the light-front spinor. In particular, the spin wave function of Eq.~(\ref{eq:15}) is obtained from the transformation of the non-relativistic spin wave function with zero orbital angular momentum component. The relativistic spin effects are immediately evident in the presence of the spin-flip term
$i\boldsymbol{\sigma}\cdot(\hat{\boldsymbol{z}}\times\boldsymbol{k}_\perp)$ in Eq.~(\ref{eq:16}).
Such a term generates non-zero orbital angular momentum, and, as a consequence of total angular momentum conservation, total quark helicity different from the nucleon helicity. Making explicit the dependence on the quark helicities, the spin wave function of Eq.~(\ref{eq:15}) takes the following values:
\begin{eqnarray}
\label{eq:17}
\tilde \Phi_\uparrow\left(\uparrow,\uparrow,\downarrow\right)
&=&\prod_i\frac{1}{\sqrt{N(x_i,\boldsymbol{ k}_{\perp i})}}
\frac{1}{\sqrt{6}}(2a_1a_2a_3+a_1 k_2^-k_3^+ +a_2k_1^-k_3^+),
\\\label{eq:18}
\tilde \Phi_\uparrow\left(\uparrow,\downarrow,\uparrow\right)
&=&\prod_i\frac{1}{\sqrt{N(x_i,\boldsymbol{ k}_{\perp i})}}
\frac{1}{\sqrt{6}}(-a_1a_2a_3+a_3 k_1^- k_2^+-2a_1k_2^+k_3^-),
\\\label{eq:19}
\tilde \Phi_\uparrow\left(\downarrow,\uparrow,\uparrow\right)
&=&\prod_i\frac{1}{\sqrt{N(x_i,\boldsymbol{ k}_{\perp i})}}
\frac{1}{\sqrt{6}}(-a_1a_2a_3+a_3 k_1^+k_2^--2a_2k_1^+k_3^-),
\\\label{eq:20}
\tilde \Phi_\uparrow\left(\uparrow,\downarrow,\downarrow\right)
&=&\prod_i\frac{1}{\sqrt{N(x_i,\boldsymbol{ k}_{\perp i})}}
\frac{1}{\sqrt{6}}(a_1a_2k_3^+-  k_1^- k_2^+k_3^+-2a_1 a_3k_2^+),
\\\label{eq:21}
\tilde \Phi_\uparrow\left(\downarrow,\uparrow,\downarrow\right)
&=&\prod_i\frac{1}{\sqrt{N(x_i,\boldsymbol{ k}_{\perp i})}}
\frac{1}{\sqrt{6}}(-k_1^+k_2^-k_3^++a_1a_2  k_3^+ -2a_2 a_3k_1^+),
\\\label{eq:21a}
\tilde \Phi_\uparrow\left(\downarrow, \downarrow, \uparrow\right)
&=&\prod_i\frac{1}{\sqrt{N(x_i,\boldsymbol{ k}_{\perp i})}}
\frac{1}{\sqrt{6}}(a_2 a_3 k_1^+ +a_1 a_3k_2^+ +2k_1^+k_2^+k_3^-),
\\\label{eq:22}
\tilde \Phi_\uparrow\left(\uparrow,\uparrow,\uparrow\right)
&=&\prod_i\frac{1}{\sqrt{N(x_i,\boldsymbol{ k}_{\perp i})}}
\frac{1}{\sqrt{6}}(-a_1a_3k_2^- - a_2a_3 k_1^-+2a_1a_2k_3^-),
\\\label{eq:23}
\tilde \Phi_\uparrow\left(\downarrow,\downarrow,\downarrow\right)
&=&\prod_i\frac{1}{\sqrt{N(x_i,\boldsymbol{ k}_{\perp i})}}
\frac{1}{\sqrt{6}}(-a_2k_1^+k_3^+ - a_1 k_2^+k_3^++2a_3k_1^+k_2^+),
\end{eqnarray}
where $a_i=(m+x_i M_0)$, and $N(x_i,\boldsymbol{ k}_{\perp i})=
[(m+x_i M_0)^2+ \boldsymbol{ k}^2_{\perp,i}]$.

Taking into account the quark-helicity dependence in Eqs.~(\ref{eq:17})-(\ref{eq:23}), the nucleon state can be mapped out into the different angular momentum components of Eq.~(\ref{eq:1}). After straightforward algebra, one finds the following representation for the nucleon amplitudes in the light-cone CQM
\begin{eqnarray}
\psi^{(1)}(1,2,3)&=&\tilde \psi(\{x_i,\boldsymbol{ k}_{\perp i}\})\nn\\
&&\times
\prod_i\frac{1}{\sqrt{N(x_i,\boldsymbol{ k}_{\perp i})}}
\frac{1}{\sqrt{3}}(
-a_1 a_2 a_3
+a_3 \boldsymbol{ k}_{1\perp}\cdot \boldsymbol{ k}_{2\perp}
+2a_1 \boldsymbol{ k}_{1\perp}\cdot \boldsymbol{ k}_{2\perp}
+2a_1 \boldsymbol{ k}_{2\perp}^2),\nn\\
&&\label{eq:24}\\
\psi^{(2)}(1,2,3)&=&\tilde \psi(\{x_i,\boldsymbol{ k}_{\perp i}\})
\prod_i\frac{1}{\sqrt{N(x_i,\boldsymbol{ k}_{\perp i})}}\frac{1}{\sqrt{3}}
(a_3 + 2 a_1),
\label{eq:25}\\
\psi^{(3)}(1,2,3)&=&-\tilde \psi(\{x_i,\boldsymbol{ k}_{\perp i}\})
\prod_i\frac{1}{\sqrt{N(x_i,\boldsymbol{ k}_{\perp i})}}
\frac{1}{\sqrt{3}}(a_1 a_2 + \boldsymbol{ k}_{2\perp}^2),
\label{eq:26}\\
\psi^{(4)}(1,2,3)&=&-\tilde \psi(\{x_i,\boldsymbol{ k}_{\perp i}\})
\prod_i\frac{1}{\sqrt{N(x_i,\boldsymbol{ k}_{\perp i})}}
\frac{1}{\sqrt{3}}(a_1 a_2 +  2a_3 a_1-\boldsymbol{ k}_{1\perp}^2 -2 
\boldsymbol{ k}_{1\perp}\cdot \boldsymbol{ k}_{2\perp}),
\label{eq:27}\\
\psi^{(5)}(1,2,3)&=&\tilde \psi(\{x_i,\boldsymbol{ k}_{\perp i}\})
\prod_i\frac{1}{\sqrt{N(x_i,\boldsymbol{ k}_{\perp i})}}
\frac{1}{\sqrt{3}}(a_1 a_3),
\label{eq:28}\\
\psi^{(6)}(1,2,3)&=&\tilde \psi(\{x_i,\boldsymbol{ k}_{\perp i}\})
\prod_i\frac{1}{\sqrt{N(x_i,\boldsymbol{ k}_{\perp i})}}
\frac{1}{\sqrt{3}} a_2.
\label{eq:29}
\end{eqnarray}
The results in Eqs.~(\ref{eq:24})-(\ref{eq:29}) follow from the spin and orbital angular momentum structure generated from the Melosh rotations, and are independent on the functional form of the momentum dependent wave function which we assumed symmetric under permutation of any quark pairs. 


\section{TMDs in a light-cone constituent quark model}
\label{sect:tmdlc}

General expressions of the TMDs can be derived under the assumption that the light-cone Fock expansion can be truncated to just the three valence quark contribution so that the nucleon wave function can be expressed in terms of six amplitudes. In terms of matrix element between proton states with different orbital angular momentum components, TMDs are given by the following expressions
\begin{eqnarray}
f_1^q(x, \boldsymbol{k}^2_\perp)&=&
^{l_z=0}\bra{P \uparrow}\sum_\lambda q^\dagger_\lambda q_\lambda\ket{P\uparrow}^{l_z=0}
+\ 
^{l_z=1}\bra{P\uparrow}\sum_\lambda q^\dagger_\lambda q_\lambda\ket{P\uparrow}^{l_z=1}\nonumber\\
&&\quad +\ 
^{l_z=-1}\bra{P\uparrow}\sum_\lambda q^\dagger_\lambda q_\lambda\ket{P\uparrow}^{l_z=-1}
+\
^{l_z=2}\bra{P \uparrow}\sum_\lambda q^\dagger_\lambda q_\lambda\ket{P\uparrow}^{l_z=2},\label{eq:f1_me}
\\
g_{1L}^q(x,\boldsymbol{k}^2_\perp)&=&
^{l_z=0}\bra{P \uparrow} \mathcal{O}_g\ket{P\uparrow}^{l_z=0}
+\ 
^{l_z=1}\bra{P\uparrow}\mathcal{O}_g\ket{P\uparrow}^{l_z=1}\nonumber\\
&&\quad +\ 
^{l_z=-1}\bra{P\uparrow}\mathcal{O}_g\ket{P\uparrow}^{l_z=-1}
+\ 
^{l_z=2}\bra{P \uparrow}\mathcal{O}_g\ket{P\uparrow}^{l_z=2},
\label{eq:g1L_me}
\\
h_1^q(x,\boldsymbol{k}^2_\perp)&=&
{\rm Re}[
^{l_z=0}\bra{P\downarrow}q^\dagger_\downarrow q_\uparrow\ket{P\uparrow}^{l_z=0}]
+\ 
2{\rm Re}[
^{l_z=-1}\bra{P\uparrow}q^\dagger_\downarrow q_\uparrow\ket{P\downarrow}^{l_z=-1}],\label{eq:h1_me}
\\
h_{1T}^{\perp \,q}(x,\boldsymbol{k}^2_\perp)&=&
-{\rm Re}[
^{l_z=1}\bra{P\uparrow}q^\dagger_\downarrow q_\uparrow\ket{P\downarrow}^{l_z=-1}]
-
2{\rm Re}[
^{l_z=0}\bra{P\uparrow}q^\dagger_\downarrow q_\uparrow\ket{P\downarrow}^{l_z=-2}],\label{eq:h1T_me}
\\
h_{1L}^{\perp \,q}(x,\boldsymbol{k}^2_\perp)&=&
\frac{2M}{k^2_\perp}
\left(
{\rm Re}[
^{l_z=1}\bra{P\uparrow}q^\dagger_\downarrow q_\uparrow \ket{P\uparrow}^{l_z=0}]
-
k^y{\rm Im}[
^{l_z=1}\bra{P\uparrow}q^\dagger_\downarrow q_\uparrow \ket{P\uparrow}^{l_z=0}]
\right.
\nonumber\\
&&\quad\quad+k^x
{\rm Re}[
^{l_z=0}\bra{P\uparrow}q^\dagger_\downarrow q_\uparrow 
\ket{P\uparrow}^{l_z=-1}]
-k^y
{\rm Im}[
^{l_z=0}\bra{P\uparrow}q^\dagger_\downarrow q_\uparrow 
\ket{P\uparrow}^{l_z=-1}]\nonumber\\
&&\left.\quad\quad+k^x
{\rm Re}[
^{l_z=2}\bra{P\uparrow}q^\dagger_\downarrow q_\uparrow 
\ket{P\uparrow}^{l_z=1}]
-k^y
{\rm Im}[
^{l_z=2}\bra{P\uparrow}q^\dagger_\downarrow q_\uparrow 
\ket{P\uparrow}^{l_z=1}]\right),
\label{eq:h1L_me}
\\
g_{1T}^{\,q}(x,\boldsymbol{k}^2_\perp)&=&
\frac{2M}{k^2_\perp}
\left(
k^x{\rm Re}[
^{l_z=0}\bra{P\uparrow}\mathcal{O}_g\ket{P\downarrow}^{l_z=-1}]
+
k^y{\rm Im}[
^{l_z=0}\bra{P\uparrow}\mathcal{O}_g\ket{P\downarrow}^{l_z=-1}]
\right.
\nonumber\\
&&\left.\quad\quad+k^x
{\rm Re}[
^{l_z=-2}\bra{P\downarrow}\mathcal{O}_g\ket{P\uparrow}^{l_z=-1}]
+k^y
{\rm Im}[
^{l_z=-2}\bra{P\downarrow}\mathcal{O}_g\ket{P\uparrow}^{l_z=-1}]\right),
\label{eq:g1T_me}
\end{eqnarray}
where $\mathcal{O}_g=\sum_\lambda (-1)^{1/2-\lambda}q^\dagger_\lambda q_\lambda$.

The unpolarized TMD $f_1^q$, the helicity TMD $g_{1L}^q$, and the transversity TMD 
$h_{1}^q$ in Eqs.~(\ref{eq:f1_me})-(\ref{eq:h1_me}) involve matrix elements which are all diagonal in the orbital angular momentum, but probe different transverse momentum and helicity correlations of the quarks inside in the nucleon.  In particular, $f^q_1$ is defined in terms of the momentum density operator,  $g^q_{1L}$ is sensitive to the difference of right and left quark-helicity. Viceversa, $h_1^q$ involves a chiral-odd operator with a quark-helicity flip compensated by a flip of the nucleon helicity in the same direction. The same chiral-odd operator enters the definition of $h_{1T}^{\perp q}$
and $h_{1L}^{\perp q}$ in Eqs.(\ref{eq:h1T_me}) and (\ref{eq:h1L_me}), respectively. In the case of $h_{1T}^{\perp q}$ the nucleon helicity flips in the direction opposite to the quark helicity, with a mismatch of the orbital angular momentum between the initial and final nucleon state of $\Delta l_z=2$, whereas $h_{1L}^{\perp q}$ is diagonal in the nucleon helicity, with the quark-helicity flip inducing a change by one unit in the orbital angular momentum of the initial and final nucleon state. Finally,  $g_{1T}^{q}$ is defined in terms of the same helicity operator which enters the definition of $g_{1L}^q$, but this time the nucleon helicity flips, with a transfer of orbital angular momentum by one unit.

If one inserts in Eqs.~(\ref{eq:f1_me})-(\ref{eq:g1T_me}) the three-quark light-cone amplitudes introduced in Eqs.~(17)-(19) and (22)-(25) of Section~\ref{sect:ampli}, one obtains the general expressions of the TMDs collected in Appendix A. These formulas can be worked out in terms of the explicit representation of the light-cone amplitudes obtained in the light-cone CQM of Section~\ref{sect:lcwf}. As we neglected gluon degrees of freedom,  the amplitudes in the light-cone CQM are pure real functions and lead to the following results for the T-even TMDs
\begin{eqnarray}
\label{eq:f1}
f^q_1(x,\boldsymbol{k}^2_\perp)&=&
N^q
\int d[1] d[2]d[3]\sqrt{x_1x_2x_3}\delta(k-k_3)
\vert \psi(\{x_i\},\{\boldsymbol{k}_{\perp,i}\})\vert^2,
\\
g^q_{1L}(x,\boldsymbol{k}^2_\perp)&=&
P^q
\int d[1] d[2]d[3]\sqrt{x_1x_2x_3}\delta(k-k_3)
\vert \psi(\{x_i\},\{\boldsymbol{k}_{\perp,i}\})\vert^2
\frac{(m+ x M_0)^2 -\boldsymbol{k}^2_{\perp}}{(m+ xM_0)^2 + \boldsymbol{k}^2_{\perp}},\nonumber\\
\label{eq:g1}&&\\
g^{q}_{1T}(x,\boldsymbol{k}^2_\perp)&=&
P^q
\int d[1] d[2]d[3]\sqrt{x_1x_2x_3}\delta(k-k_3)
\vert \psi(\{x_i\},\{\boldsymbol{k}_{\perp,i}\})\vert^2
\frac{2M(m+ xM_0)}{(m+ xM_0)^2 + \boldsymbol{k}^2_{\perp}}, \nonumber\\
&&
\label{eq:g1T}\\
h^q_1(x,\boldsymbol{k}^2_\perp)&=&
P^q
\int d[1] d[2]d[3]\sqrt{x_1x_2x_3}\delta(k-k_3)
\vert \psi(\{x_i\},\{\boldsymbol{k}_{\perp,i}\})\vert^2
\frac{(m+ xM_0)^2}{(m+ xM_0)^2 + \boldsymbol{k}^2_{\perp}},\nonumber\\
&&
\label{eq:h1}\\
h^{\perp\,q}_{1T}(x,\boldsymbol{k}^2_\perp)&=&-
P^q
\int d[1] d[2]d[3]\sqrt{x_1x_2x_3}\delta(k-k_3)
\vert \psi(\{x_i\},\{\boldsymbol{k}_{\perp,i}\})\vert^2
\frac{2M^2}{(m+ xM_0)^2 + \boldsymbol{k}^2_{\perp}},\nonumber\\
\label{eq:h1T}&&\\
h^{\perp\, q}_{1L}(x,\boldsymbol{k}^2_\perp)&=&
- P^q
\int d[1] d[2]d[3]\sqrt{x_1x_2x_3}\delta(k-k_3)
\vert \psi(\{x_i\},\{\boldsymbol{k}_{\perp,i}\})\vert^2
\frac{2M(m+ xM_0)}{(m+ xM_0)^2 + \boldsymbol{k}^2_{\perp}},\nonumber\\
&&
\label{eq:h1L}
\end{eqnarray}
where $\delta(k-k_3)= \delta(x-x_3)\delta(\boldsymbol{k}_{\perp}-\boldsymbol{k}_{\perp\,3})$, and the flavor factors $P^u=\displaystyle{\frac{4}{3}}$,  $P^d=\displaystyle{-\frac{1}{3}}$, $N^u=2$ and $N^d=1$ are dictated by SU(6) symmetry.

By inspection the above TMDs satisfy the following relations:
\begin{eqnarray}
\label{eq:61}
2h^q_1(x,\boldsymbol{k}^2_\perp)
&=&g^q_{1L}(x,\boldsymbol{k}^2_\perp)+\frac{P^q}{N^q}f^q_1(x,\boldsymbol{k}^2_\perp),\\
\frac{P^q}{N^q}f^q_1(x,\boldsymbol{k}^2_\perp)
&=&h_1^q(x,\boldsymbol{k}^2_\perp) -\frac{\boldsymbol{k}^2_\perp}{2M^2}h_{1T}^{\perp \,q}(x,\boldsymbol{k}^2_\perp),
\label{eq:61a}\\
h_{1L}^{\perp q}(x,\boldsymbol{k}^2_\perp)
&=&-g_{1T}^q(x,\boldsymbol{k}^2_\perp).
\label{eq:61b}
\end{eqnarray}
Eq.~(\ref{eq:61}) is a generalization of analogous relations discussed in Ref.~\cite{PPB,PPB07} and was also rederived  together with Eq.~(\ref{eq:61a}) in Ref.~\cite{Avakian08}. Eq.~(\ref{eq:61b}) was already found in the diquark spectator model of Ref.~\cite{Mulders3}. In QCD TMDs should be all independent of each other. The limitation to three valence quarks implies that out of the six TMDs $f_1$, $g_{1L}$, $g_{1T}$, $h_1$,  $h_{1T}^\perp$, $h_{1L}^\perp$ only three are linearly independent. A similar situation occurs with the bag model~\cite{Avakian08}. In the bag model there are only S- and P-wave components of the proton wave function, whereas here also a D-wave contributes. However, the relations (\ref{eq:61})-(\ref{eq:61b}) do not depend on the different components of orbital angular momentum. Their specific form is a consequence of the imposed SU(6) symmetry which allows us to factorize the momentum dependent wave function from the effects of the Melosh rotation acting in the spin space and producing different factors for the different TMDs, Eqs.~(\ref{eq:f1})-(\ref{eq:h1L}). In the diquark spectator model the relations (\ref{eq:61}) and (\ref{eq:61a}) hold only for the separate scalar and axial contributions, while Eq.~(\ref{eq:61b}) is verified more generally for both $u$ and $d$ flavors. 

Concerning what has been called the `pretzelosity' distribution, $h_{1T}^\perp$~\cite{Avakian08}, 
from Eqs.~(\ref{eq:f1}), (\ref{eq:g1}), and (\ref{eq:h1T}) one easily verifies 
the positivity condition~\cite{BBHM}
\begin{eqnarray}
\left|\frac{\boldsymbol{k}_\perp^2}{2M^2}\,h_{1T}^{\perp\,q}(x,\boldsymbol{k}^2_\perp)\right|
\le\frac{1}{2}\left(f_1^q(x,\boldsymbol{k}^2_\perp)-g_1^q(x,\boldsymbol{k}^2_\perp)\right)\le f_1^q(x,\boldsymbol{k}^2_\perp).
\end{eqnarray}
Furthermore, subtracting the relations (\ref{eq:61}), (\ref{eq:61a}) from each other yields
\be
\frac{\boldsymbol{k}_{\perp}^2}{2M^2}\,h^{\perp\,q}_{1T}(x,\boldsymbol{k}^2_\perp) 
= g^q_{1L}(x,\boldsymbol{k}^2_\perp) - h^q_1(x,\boldsymbol{k}^2_\perp).
\label{eq:pretz}
\ee
This result was already found in Ref.~\cite{Avakian08}. Integrating out transverse momenta and going to the non-relativistic limit where helicity and transversity distributions coincide, one finds that the first moment of $h_{1T}^\perp$ vanish identically. Thus, relation (\ref{eq:pretz}) supports the statement that $h_{1T}^\perp$ is a measure of relativistic effects. Relativity, responsible for a chiral-odd  transversity distribution differing from a chiral-even helicity distribution, exhibits the chirally odd nature of  $h_{1T}^\perp$. This is confirmed by the following relation that is also satisfied within our model:
\begin{eqnarray}
\label{eq:tmd_gpd}
h_{1T}^{(0)\perp\,q}(x)=\frac{3}{(1-x)^2}\tilde H_T^q(x,0,0),
\end{eqnarray}
where the transverse moments of $h_{1T}^{\perp\,q}$ are defined as
\be
h_{1T}^{(n)\perp q}(x)=\int d^2 k_\perp \left(\frac{\boldsymbol{k}_{\perp}^2}{2M^2}\right)^n \, h_{1T}^{\perp\,q}(x,\boldsymbol{k}^2_\perp),
\ee
and $\tilde H_T^q(x,0,0)$ is the forward limit of a chiral-odd generalized parton distribution (GPD) occurring in the case of parton and nucleon helicity flip (see, e.g., Refs.~\cite{PPB,BPreview}). Eq.~(\ref{eq:tmd_gpd}) was first found in Ref.~\cite{Meissner} to hold for the scalar diquark model and in a quark target model of the nucleon as a particular case of a general relation between the moments of TMDs and the moments of GPDs.

By integrating over $x$ the first moment of $h_{1T}^\perp$, from Eq.~(\ref{eq:pretz}) one obtains
\be
\int dx\, h_{1T}^{(1)\perp q} (x)= \Delta q - \delta q,
\ee
where $\Delta q$ and $\delta q$ are the axial and tensor charges that measure, for each flavor $q$, 
the net number of longitudinally polarized valence quarks in a longitudinally polarized nucleon
and 
the net number of transversely polarized valence quarks in a transversely polarized nucleon, 
respectively.

\section{Results}
\label{sect:result}

The full list of T-even quark TMDs was computed in Ref.~\cite{Mulders3} (see also ~\cite{Conti,Gamberg}) using the diquark spectator model with scalar and 
axial-vector diquark. The analytic form of TMDs was also derived in a quark target model~\cite{Meissner}. Here, the formalism described in the previous sections is applied in the following to a specific CQM adopting a power-law form for the momentum dependent part of the light-cone wave function~\cite{Schlumpf:94a}, i.e.
\bea
\tilde \psi(\{x_i,\boldsymbol{ k}_{\perp i}\})=
2(2\pi)^3\bigg[\frac{1}{M_0}\frac{\omega_1\omega_2\omega_3}{x_1x_2x_3}\bigg]^{1/2}
\frac{N'}{(M_0^2+\beta^2)^\gamma},
\label{eq:30}
\eea 
with $N'$  a normalization factor. In Eq.~(\ref{eq:30}), the scale $\beta$, the parameter $\gamma$ for the power-law behaviour, and the quark mass $m$ are taken from Ref.~\cite{Schlumpf:94a}, i.e. $\beta=0.607 $ GeV, $\gamma=3.4$ and $m=0.267$ GeV. According to the analysis of Ref.~\cite{Schlumpf:94b} these values lead to a very good description of many baryonic properties.

\begin{figure}[ht]
\begin{center}
\epsfig{file=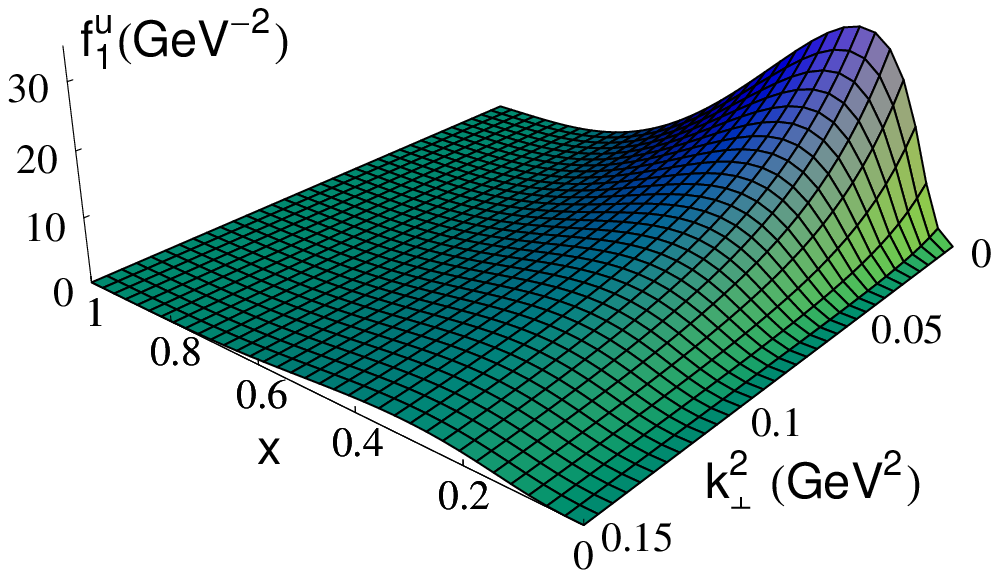,  width=8.1 truecm}
\epsfig{file=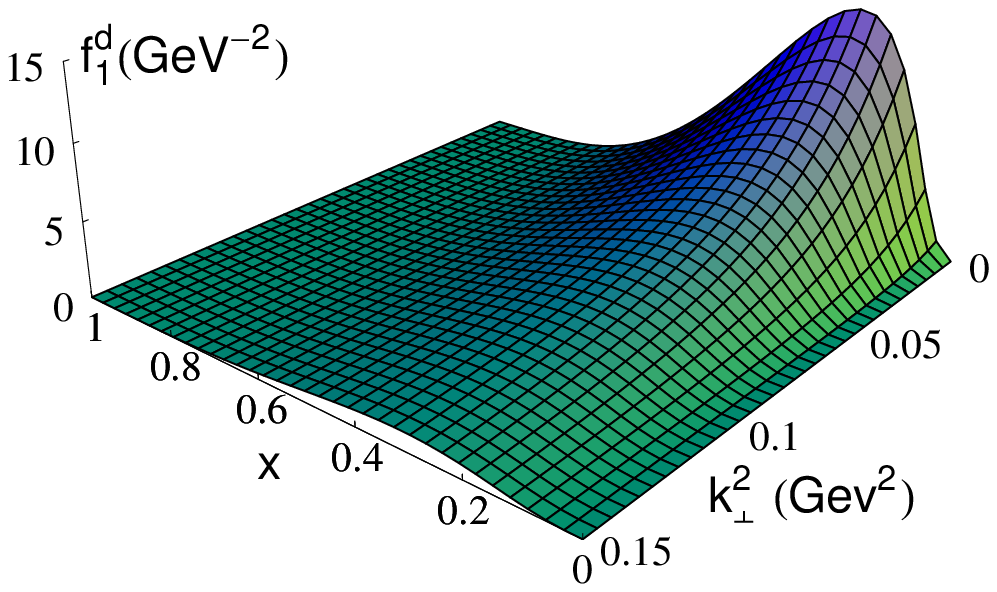,  width=8.1 truecm}
\smallskip
\epsfig{file=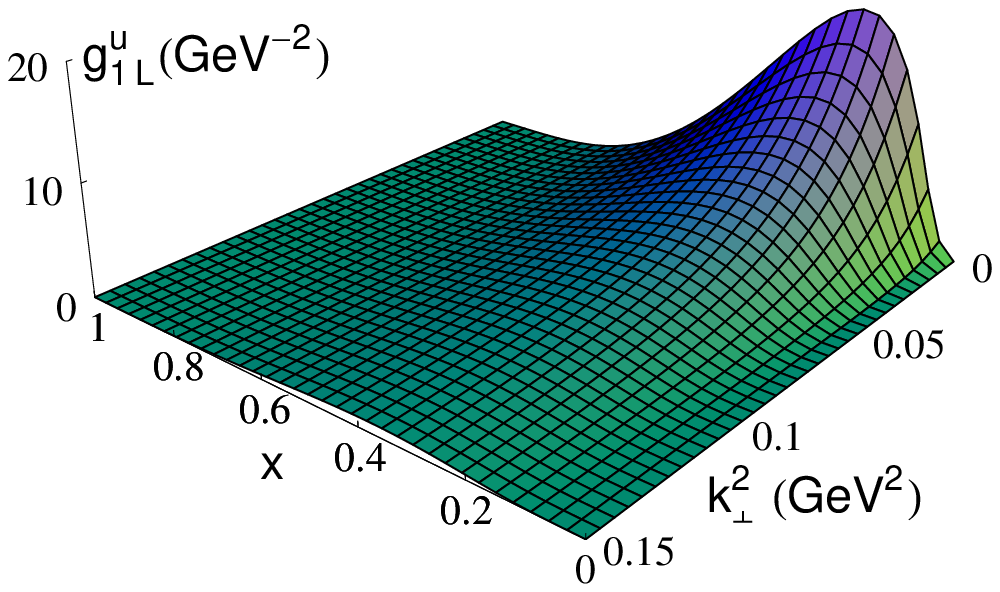,  width=8.1 truecm}
\epsfig{file=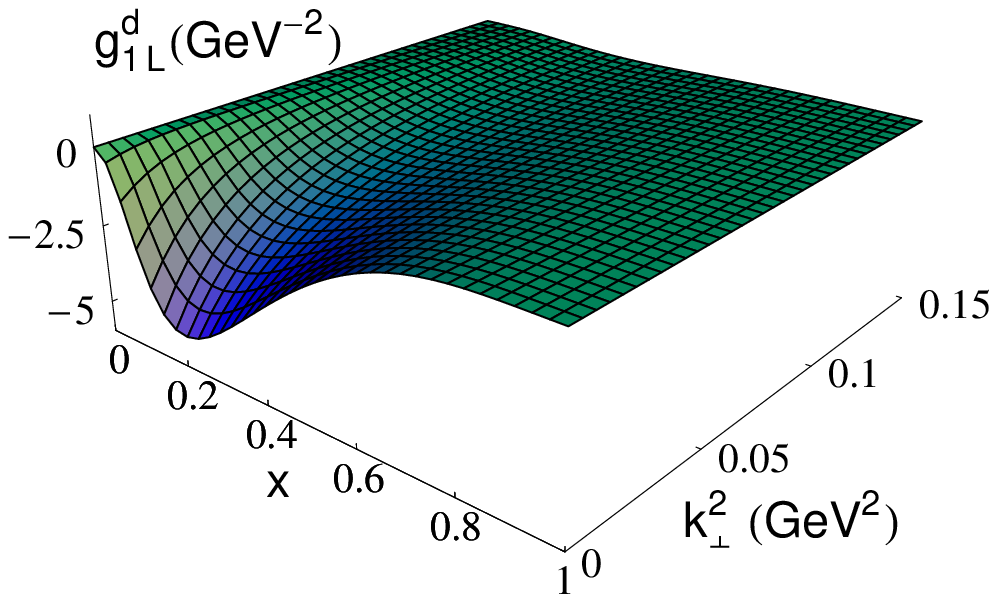,  width=8.1 truecm}
\smallskip
\epsfig{file=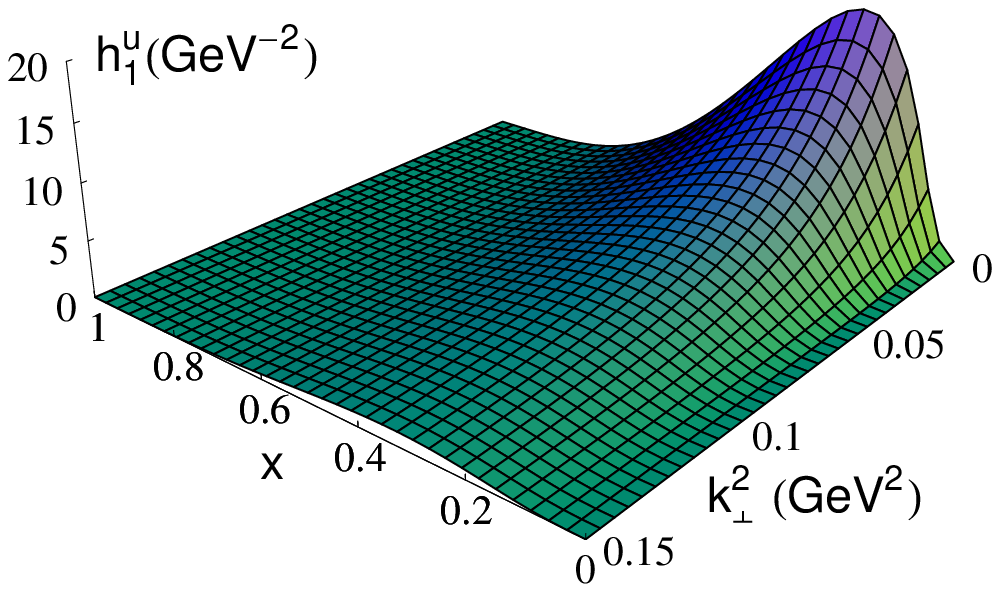,  width=8.1 truecm}
\epsfig{file=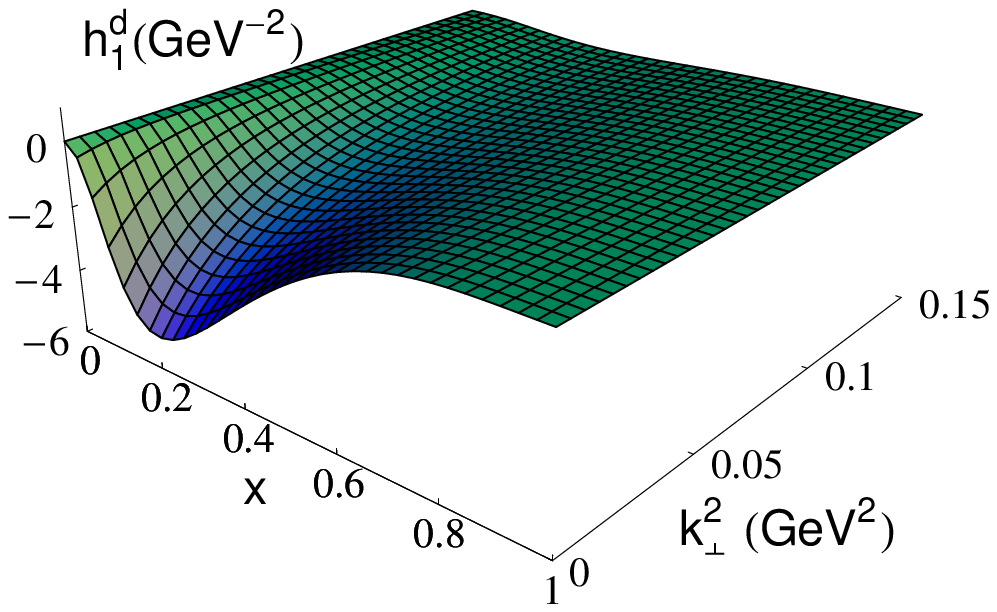,  width=8.1 truecm}
\end{center}
\caption{(Color online) The TMDs $f_1$, $g_{1L}$, $h_1$ as functions of $x$ and $\boldsymbol{k}^2_\perp$ are shown in the upper, middle and lower panels, respectively. Results for up and down quarks are given in the left (right) panels. 
}
\label{fig:fgh1}
\end{figure}

The results for $f_1^q$, $g_{1L}^q$ and $h_1^q$ are shown in Fig.~\ref{fig:fgh1}. They are consistent with those obtained in Ref.~\cite{PPB07} for the corresponding PDFs indicating that $f_1^u$ and $f_1^d$ have the same (positive) sign, whereas $g_{1L}^u $ and $h_1^u$ have opposite sign with respect to $g_{1L}^d$ and $h_1^d$, respectively. In addition, the size of the TMDs for $d$ quarks is smaller than that for $u$ quarks according to the flavor dependence of TMDs through the factor $P^q$ in Eqs.~(\ref{eq:g1})-(\ref{eq:h1L}). It is remarkable that the $\boldsymbol{k}_\perp^2$ dependence cannot be factorized as a Gaussian function as often assumed. In all distributions the maximum (minimum) is around $x= 0.2$ at $\boldsymbol{k}_\perp^2=0$ and moves to higher values of $x$ with increasing $\boldsymbol{k}_\perp^2$.

\begin{figure}[ht]
\begin{center}
\epsfig{file=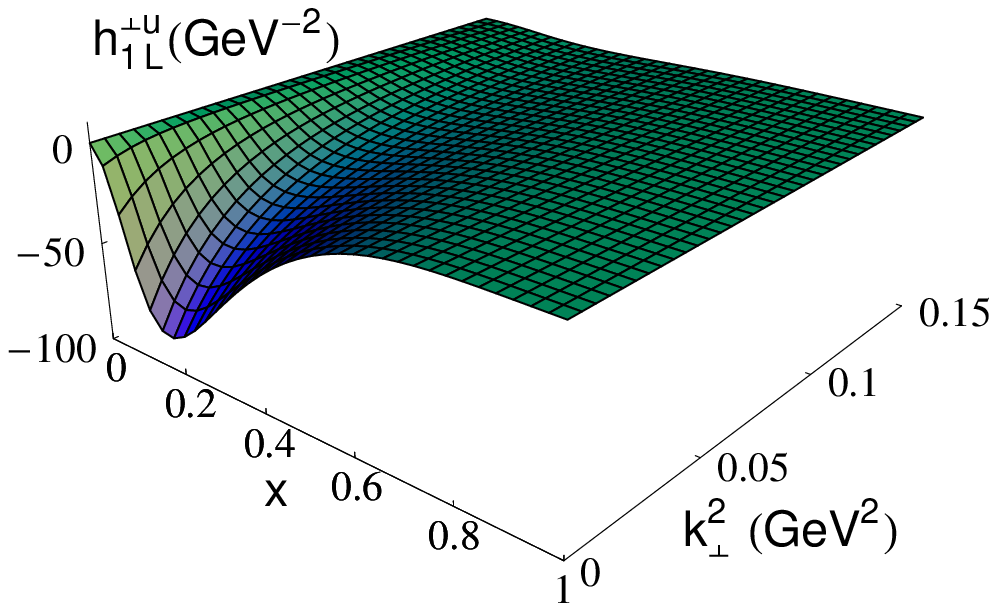,  width=8.1 truecm}
\epsfig{file=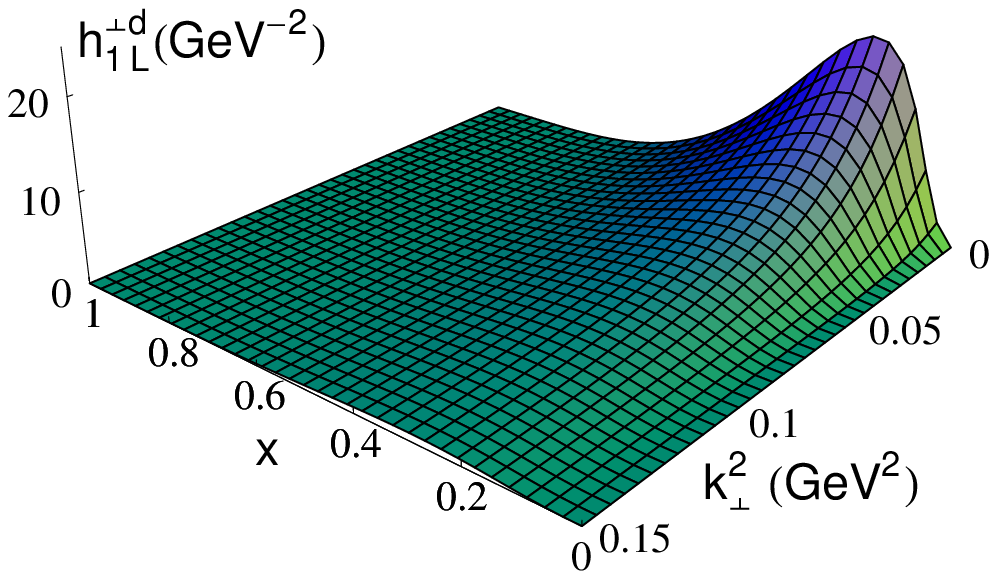,  width=8.1 truecm}
\smallskip
\epsfig{file=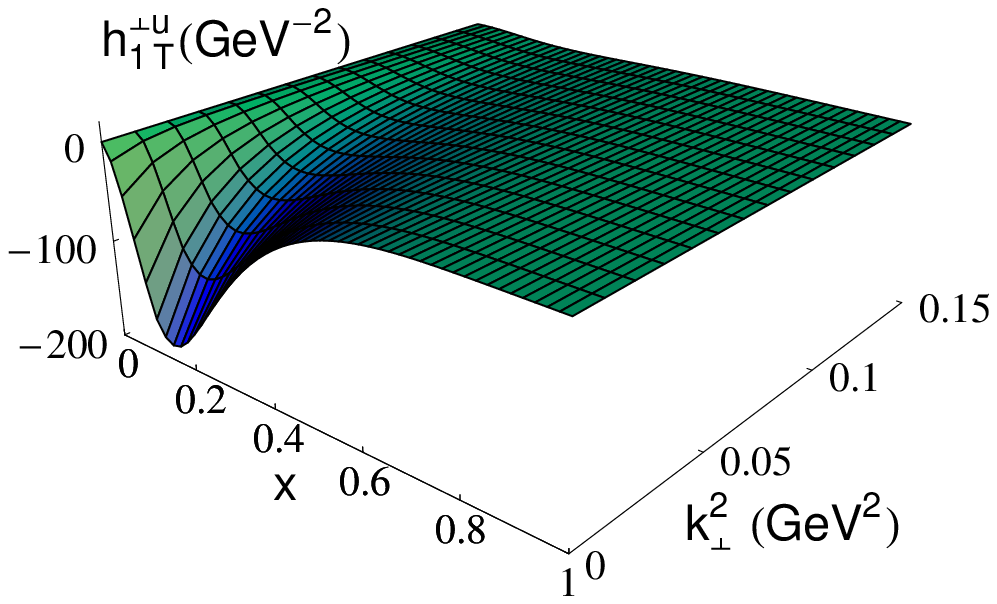,  width=8.1 truecm}
\epsfig{file=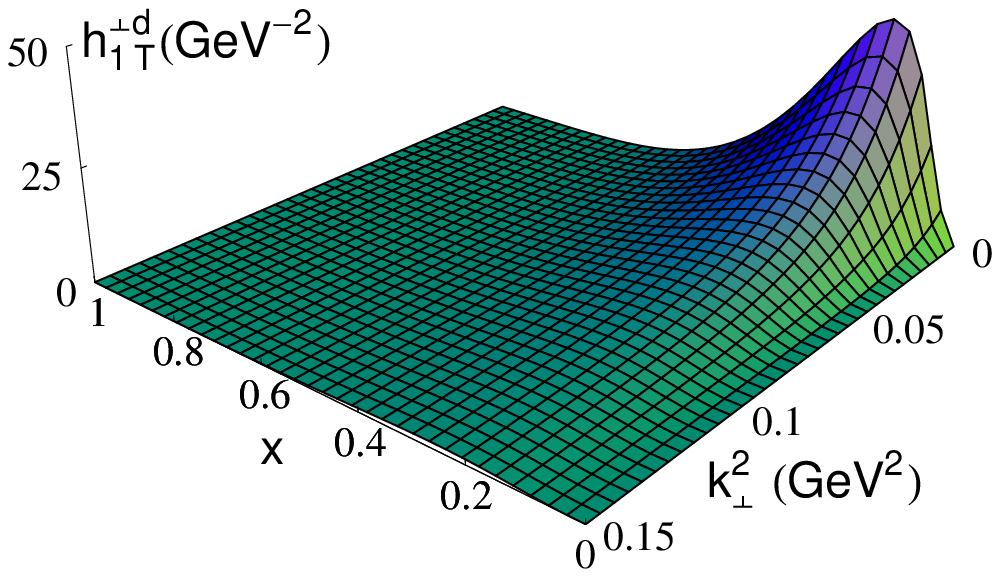,  width=8.1 truecm}
\end{center}
\caption{(Color online) The TMDs $h_{1L}$  and $h_{1T}^\perp$ as functions of $x$ and $\boldsymbol{k}^2_\perp$ are shown in the upper and lower panels, respectively. Results for up and down quarks are given in the left (right) panels. 
}
\label{fig:h1LT}
\end{figure}

\begin{figure}[ht]
\begin{center}
\epsfig{file=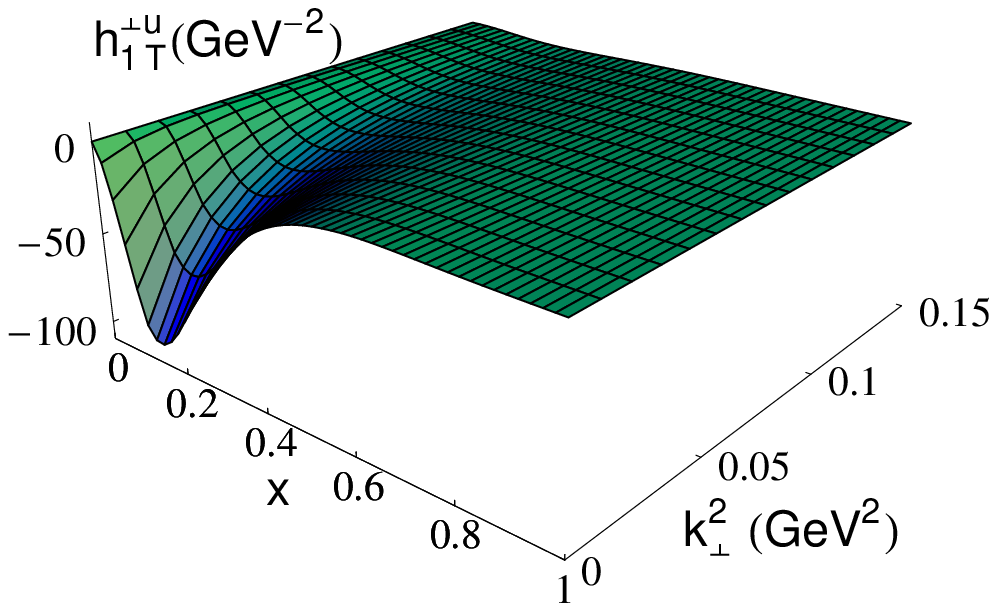,  width=8.1 truecm}
\epsfig{file=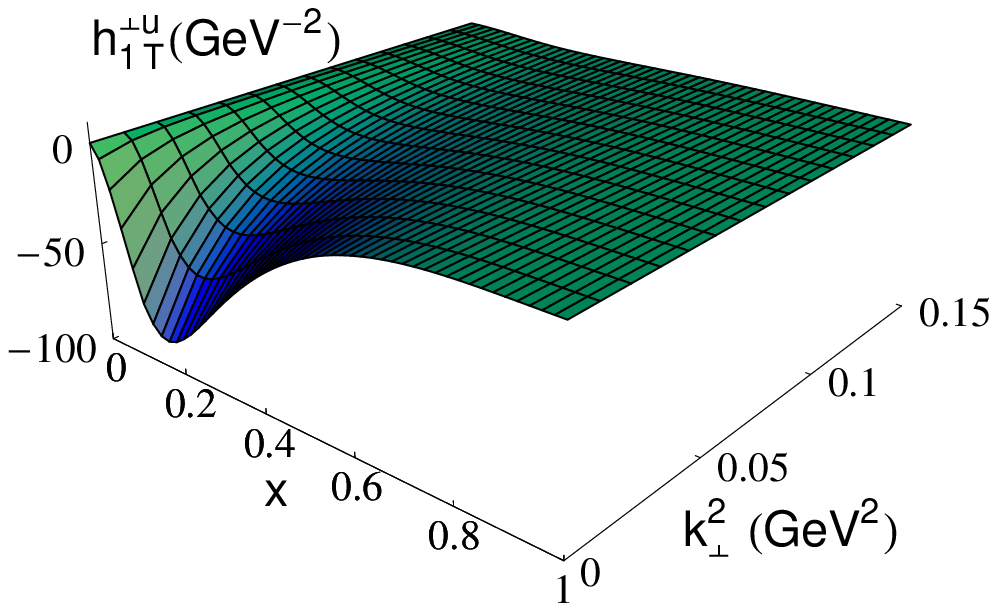,  width=8.1 truecm}
\smallskip
\epsfig{file=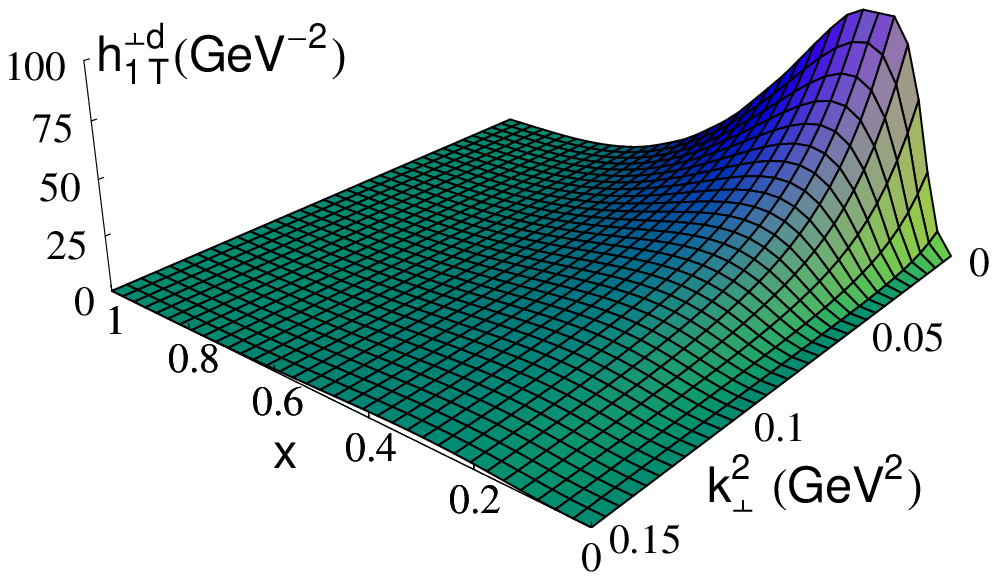,  width=8.1 truecm}
\epsfig{file=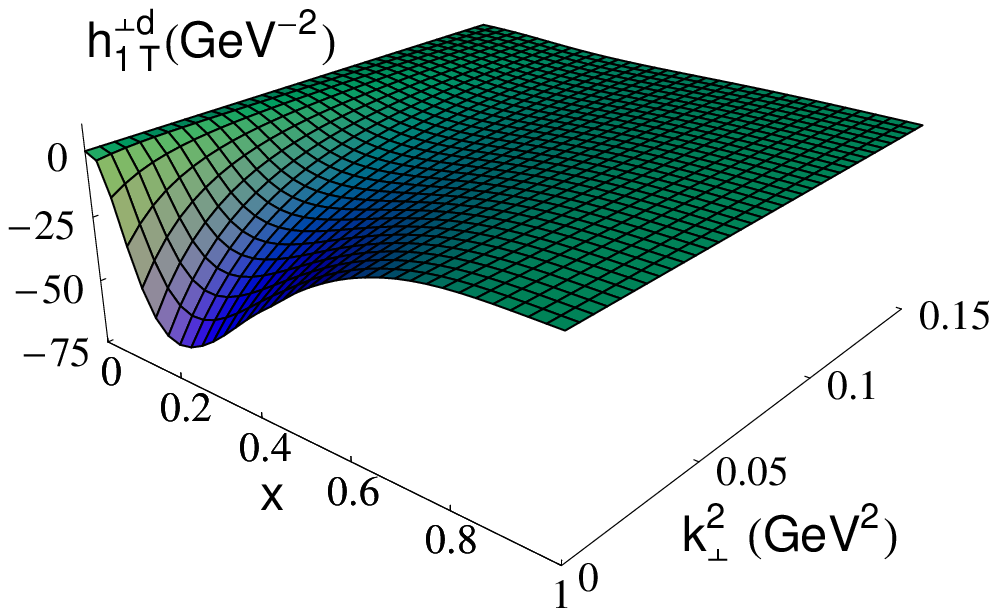,  width=8.1 truecm}
\end{center}
\caption{(Color online) The contribution to the TMD $h_{1T}^{u\perp}$ from the $L_z=\pm 1$ wave components (left panel) and  (right panel) from the $L_z=0$ and $L_z=2$ wave components as a function of $x$ and $\boldsymbol{k}^2_\perp$. Upper (lower) panels for up and down quarks.
}
\label{fig:h1T_part}
\end{figure}

The TMDs $h_{1L}^{\perp\,q}$ and $h_{1T}^{\perp\,q}$ are shown in Fig.~\ref{fig:h1LT}. The size of both $h_{1L}^{\perp\,q}$ and $h_{1T}^{\perp\,q}$ is much larger than that of $f_1^q$, $g_{1L}^q$ and $h_1^q$, a result in qualitative agreement with that obtained in the bag model~\cite{Meissner}. In particular, comparing $h_1^q$ and $h_{1L}^{\perp\,q}$ one deduces that the quark helicity flip is more favored in the case of a longitudinally polarized nucleon with a transfer of orbital angular momentum between initial and final states. The shapes of the $x$-distributions of $h_1^q$ and $h_{1L}^{\perp\,q}$ are similar, but with opposite sign. On the other hand, $h_{1T}^{\perp\,q}$ has a narrower $x$-distribution with a faster fall-off in $\boldsymbol{k}_\perp^2$. The pretzelosity $h_{1T}^{\perp\,q}$ contributes when the quark and nucleon helicity flip in opposite directions. It then requires an overlap between wave function components that differ by two units of orbital angular momentum, either a PP or an SD interference (see Eq.~(\ref{eq:h1T_me})). The different partial wave contributions to $h_{1T}^{\perp\,q}$ are plotted in Fig.~\ref{fig:h1T_part} where one may notice the importance of also including the D-wave component which is absent, e.g., in the bag model and the diquark spectator model. While in the case of $u$ quarks the PP and SD interference terms add with the same sign, in  the case of $d$ quarks they have opposite sign, indicating that the SU(6) relation between $u$ and $d$ contributions, $h_{1T}^{\perp\,u}= -4 h_{1T}^{\perp\,d}$, is valid for the total result but not for the partial wave contributions. 

\begin{figure}[ht]
\begin{center}
\epsfig{file=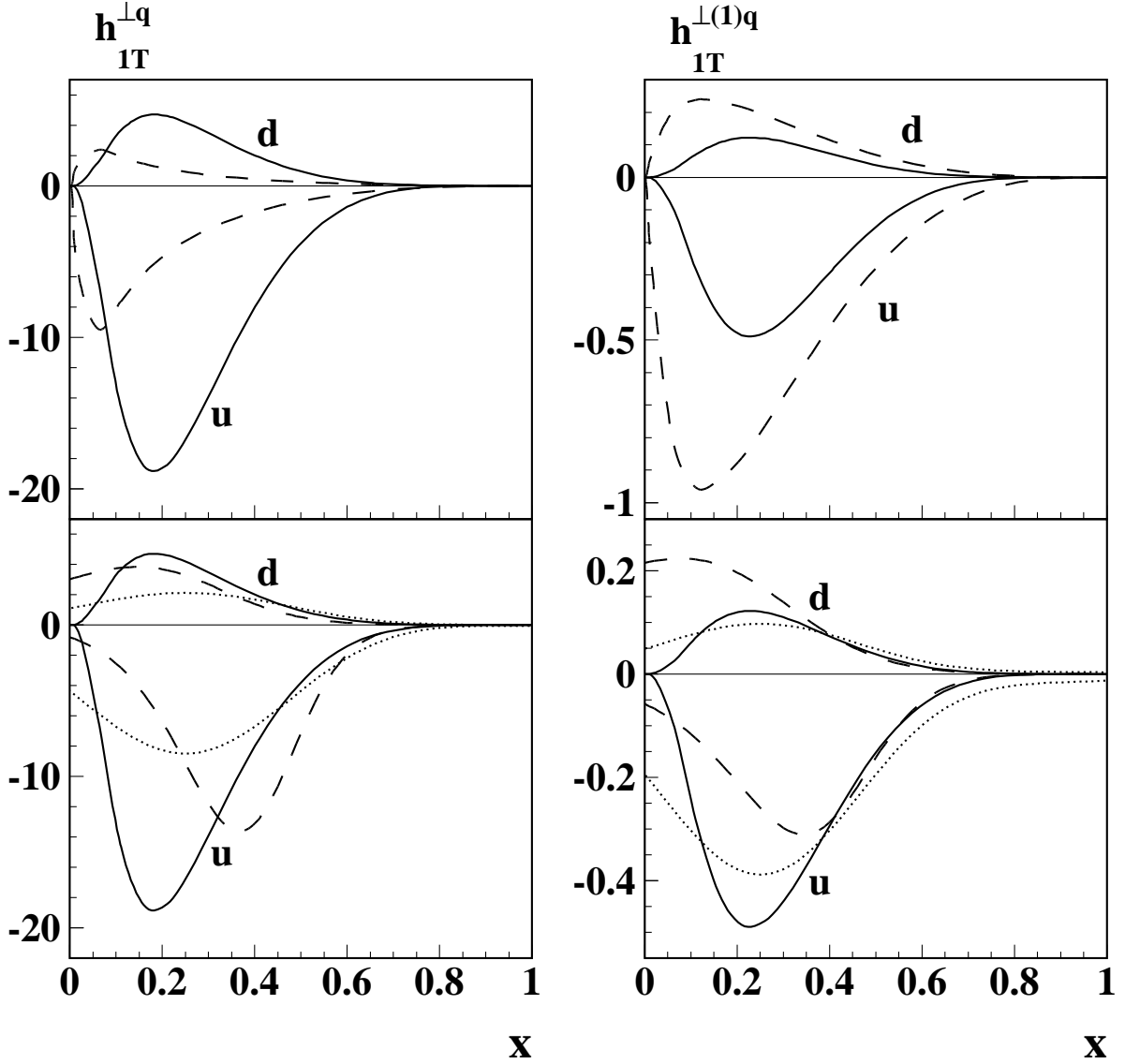,  width=16 truecm}
\end{center}
\caption{The parton distribution $h_{1T}^{\perp q}(x)$  (left panels) and the transverse moment $h_{1T}^{(1)\perp q}(x)$ (right panels).
Solid curves: results from the light-cone CQM model with the momentum wave function of Ref.~\cite{Schlumpf:94a}.
Dashed curves in the upper panels:  results from the light-cone CQM model with the momentum wave function in the hypercentral model of Ref.~\cite{Faccioli,Giannini}.
Dashed curves in the lower panels: results from the spectator model of Ref.~\cite{Mulders3}. Dotted curves: results from the bag model.
}
\label{fig:h1T_int}
\end{figure}

The results presented in Figs.~\ref{fig:fgh1} and \ref{fig:h1LT} are qualitatively similar also to those obtained with the diquark spectator model~\cite{Mulders3}. With respect to the results obtained in Ref.~\cite{Mulders3} with scalar and axial masses $M_R=0.6$ and $0.8$ GeV, respectively, and with a cut-off $\Lambda=0.5$ GeV, TMDs calculated with the light-cone CQM are peaked at smaller values of $x$ with broader $x$ and $\boldsymbol{k}_\perp^2$ distributions. As a consequence, the transverse moments of the pretzelosity distribution may be rather different in different models, as can be seen in Fig.~\ref{fig:h1T_int} where the light-cone CQM results with the momentum wave function~(\ref{eq:30}) are compared with similar calculations with another momentum wave function derived within the hypercentral model~\cite{Faccioli,Giannini} and with results of the bag model and the spectator model of Ref.~\cite{Mulders3}. This sensitivity to the adopted model suggests that new data could give useful insights to model the momentum dependence of the nucleon wave function.

According to Eq.~(\ref{eq:61b}), $g_{1T}^q$ is just the opposite of $h_{1L}^{\perp\,q}$. Both TMDS $g_{1T}^q$ and $h_{1L}^{\perp\,q}$ involve matrix elements between states that differ by one unit of angular momentum and one finds here $h_{1L}^{\perp\,u}<0<h_{1L}^{\perp\,d}$ and $g_{1T}^u>0>g_{1T}^d$, in agreement with some expectation~\cite{Burkardt07a}. On the contrary, in our model $h_{1T}^{\perp\,u}$ and $h_{1T}^{\perp\,d}$ are obtained with the reversed sign predicted by qualitative arguments in Ref.~\cite{Burkardt07a} but in agreement with the result within the bag model~\cite{Meissner} and the diquark spectator model~\cite{Mulders3}. 

\begin{figure}[ht]
\begin{center}
\epsfig{file=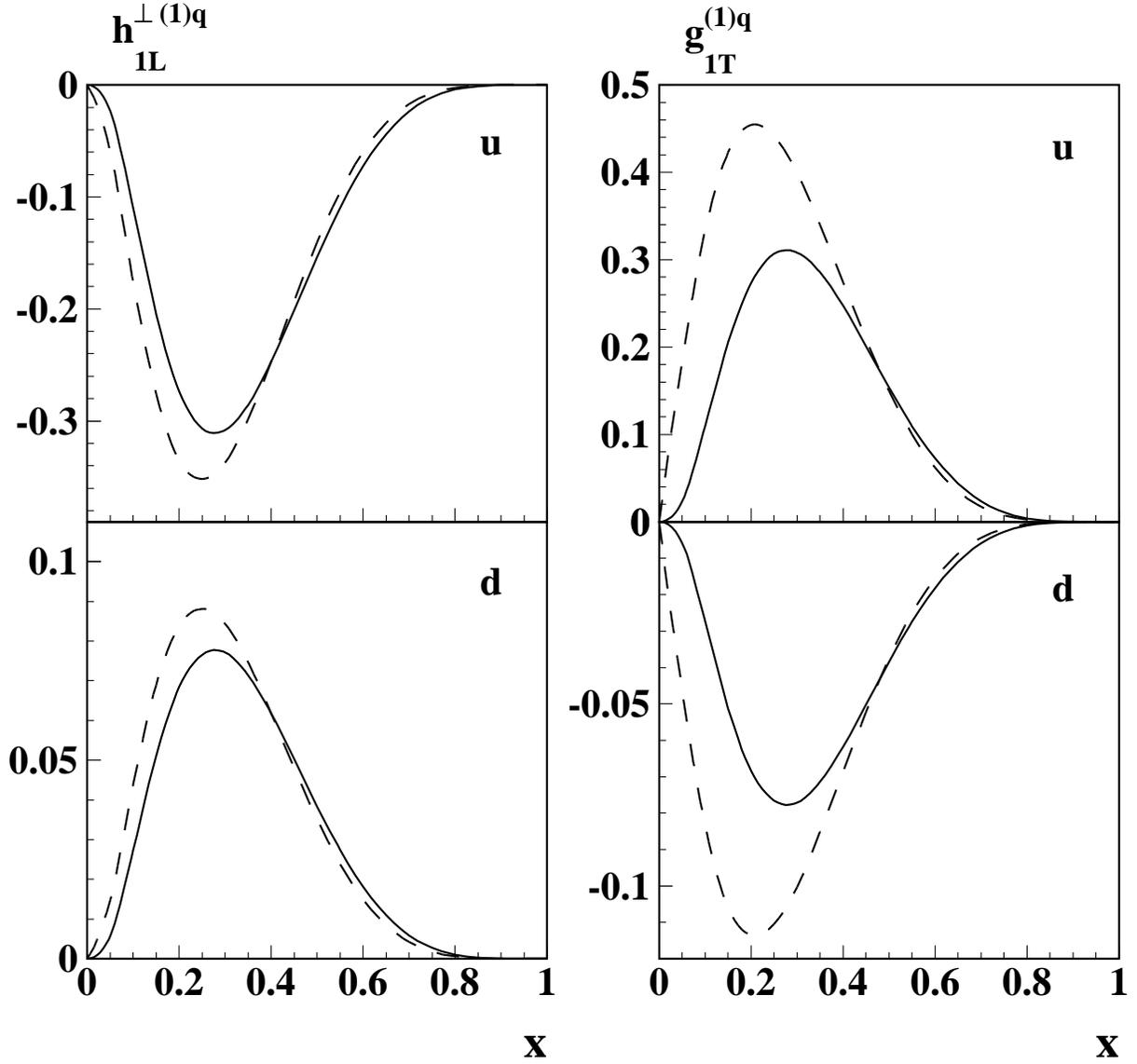,  width=16 truecm}
\end{center}
\caption{The transverse moments $h_{1L}^{\perp(1)\,q}$ (left panels) and $g_{1T}^{(1)\,q}$ (right panels) as functions of $x$. Upper (lower) panels for up and down quarks. Solid curves refer to the result obtained with the light-cone CQM model; dashed curves refer to the Wandzura-Wilczek-type approximation.
}
\label{fig:ww}
\end{figure}

In Ref.~\cite{Avakian08a} approximate relations among TMDs were studied. Taking advantage of the QCD equation of motion and neglecting twist-three TMDs, Wandzura-Wilczek-type approximations were proposed for the transverse moments of $g_{1T}^q$ and $h_{1L}^{\perp\,q}$, i.e.
\bea
g_{1T}^{(1)\,q} (x) &\approx& x\int_x^1\frac{dy}{y}\, g_1^q(y), 
\label{eq:g1t} \\
h_{1L}^{\perp(1)\,q}(x) & \approx& - x^2\int_x^1 \frac{dy}{y^2}\, h_1^q(y).
\label{eq:h1lperp}
\eea
According to Eq.~(\ref{eq:61b}), $g_{1T}^{(1)\,q}$ and $h_{1L}^{\perp(1)\,q}$ should be equal and with opposite sign in the present model, whereas the helicity and transversity distributions, $g_1^q$ and $h_1^q$ respectively, are rather different~\cite{PPB07}. 
As a consequence, in Fig.~\ref{fig:ww} one may appreciate how good is the Wandzura-Wilczek-type 
approximation when considering an SU(6) symmetric 
model as the one adopted here. In fact, the approximation of neglecting 
twist-three contributions works better for $h_{1L}^{\perp(1)\,q}$ than for $g_{1T}^{(1)\,q}$.
In any case, the model results support that the estimates for
     spin observables in SIDIS made in Refs.~\cite{Avakian08a,Kotzinian:2006dw} on the basis
     of the approximations (66) and (67) have a useful accuracy.

The presence of a significant orbital angular momentum component is suggesting a nonspherical shape of the nucleon. According to Ref.~\cite{Miller08} there is an infinite variety of obtainable shapes depending on the contribution of the pretzelosity $h_{1T}^\perp$. They can be found by looking at a suitable spin-dependent quark density,  $\hat \rho_{{\rm REL}\,T}$, in a nucleon state polarized in the transverse direction $\boldsymbol{S}_T$ either parallel or antiparallel to a given direction $\boldsymbol{n}$. The transverse shapes of the nucleon are then derived from the following relation:
\be
\frac{\hat \rho_{{\rm REL}\,T}(\boldsymbol{k}_\perp,\boldsymbol{n})/M}{\tilde f_1(\boldsymbol{k}_\perp^2)}
=
1 + \frac{\tilde h _1(\boldsymbol{k}_\perp^2)}{\tilde f_1(\boldsymbol{k}_\perp^2)}\cos\phi_n + \frac{\boldsymbol{k}_\perp^2}{2M^2}\cos(2\phi-\phi_n)\frac{\tilde h_{1T}^\perp(\boldsymbol{k}_\perp^2)}{\tilde f_1(\boldsymbol{k}_\perp^2)}, 
\label{eq:rhot}
\ee
where $\phi$ is the angle between $\boldsymbol{k}_\perp$ and $\boldsymbol{S}_T$ and $\phi_n$ is the angle between $\boldsymbol{n}$ and $\boldsymbol{S}_T$. A tilde is placed over a given quantity to define the $x$-integrated result, e.g.,
\be
\tilde f_1(\boldsymbol{k}_\perp^2) = \int dx\, f_1(x, \boldsymbol{k}_\perp^2).
\ee

\begin{figure}[ht]
\begin{center}
\epsfig{file=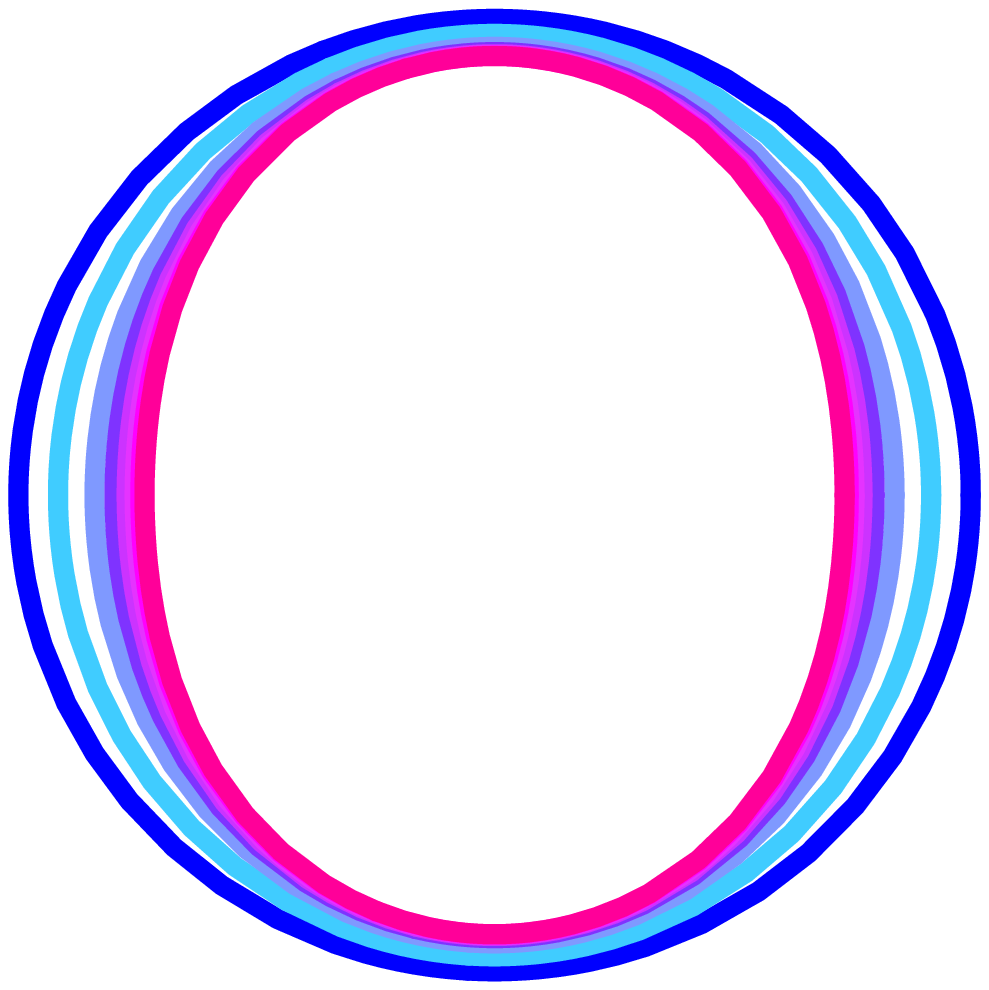,  width=7.5 truecm}
\hspace{1 truecm}
\epsfig{file=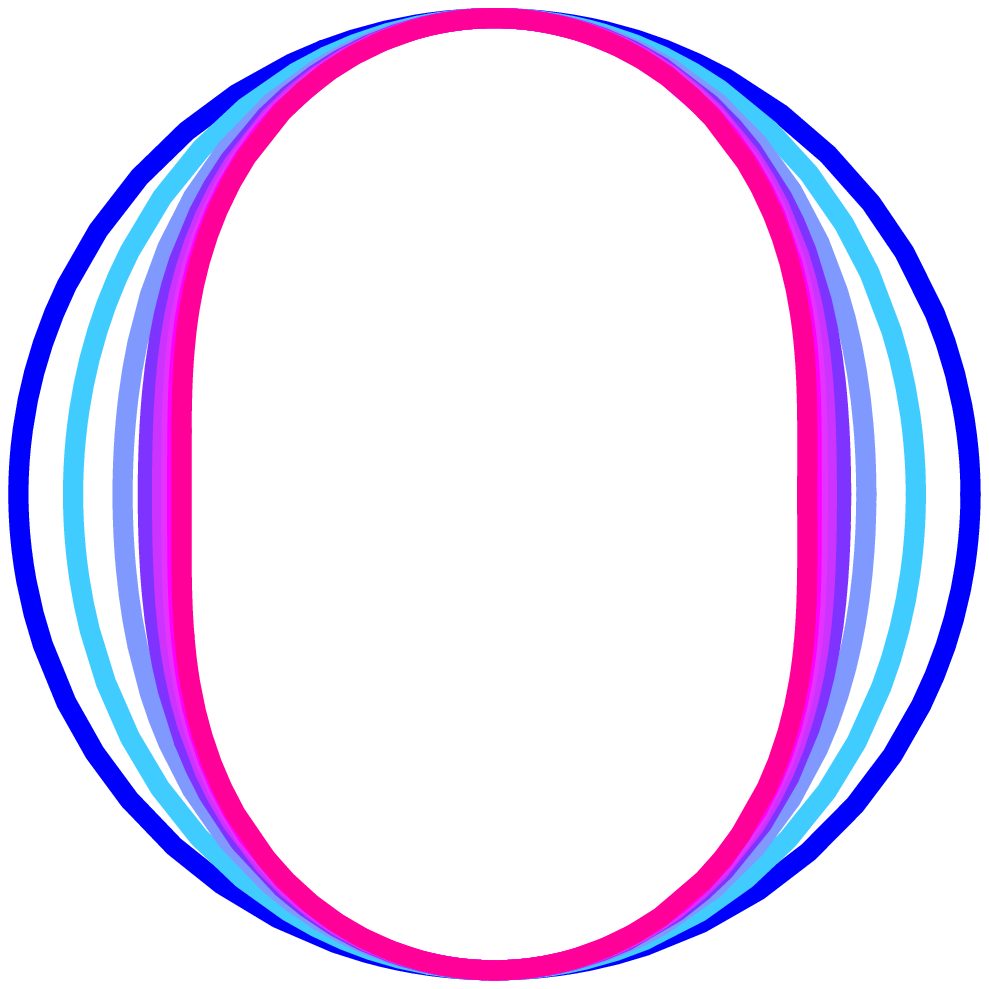,  width=7.5 truecm}
\end{center}
\caption{(Color online) Transverse shape of the proton,
$\hat \rho_{{\rm REL}\,T}(\boldsymbol{k}_\perp,\boldsymbol{n})/\tilde f_1(\boldsymbol{k}_\perp^2)$,
assuming a struck $u$ quark. The horizontal axis is the direction of $\boldsymbol{S}_\perp$ and 
$\boldsymbol{n}=\hat{\boldsymbol{S}}_\perp$,  $\phi_n=0$. The shapes vary from the outer circle to the internal line as $k_\perp$ is increased from 0 to 2.0 GeV in steps of 0.25 GeV. The left figure is the results when only the contribution from the P waves to $\tilde h_{1T}^\perp$ is taken into account, and the right picture shows the total results when also the S and D waves are included.  
}
\label{fig:density_plus_up}
\end{figure}

\begin{figure}[ht]
\begin{center}
\epsfig{file=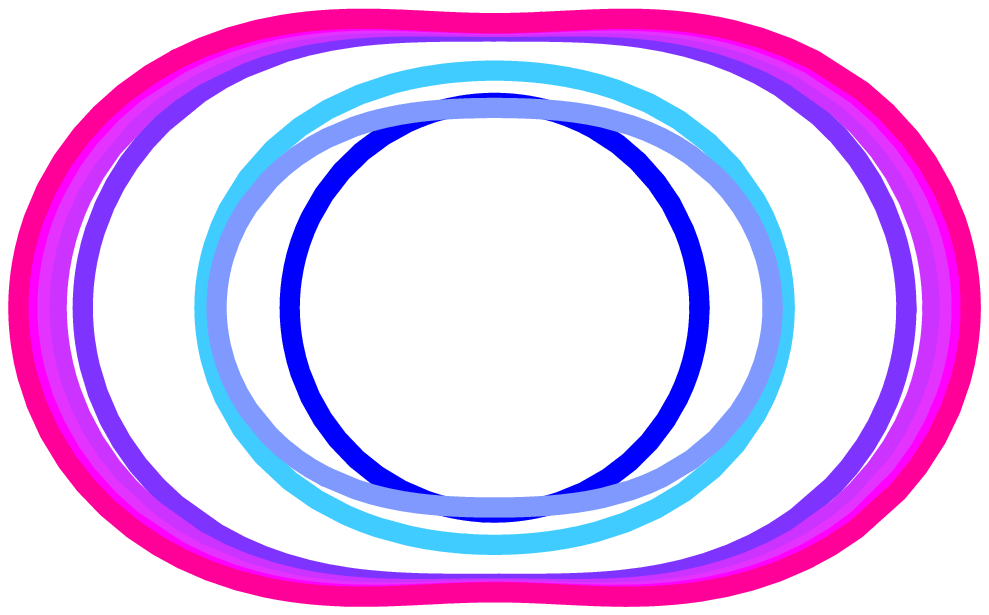,  width=7.5 truecm}
\hspace{1 truecm}
\epsfig{file=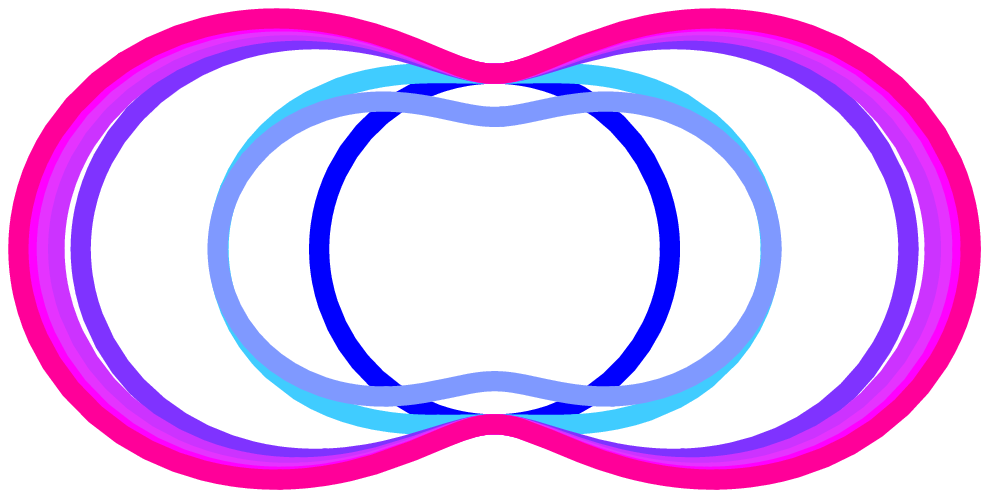,  width=7.5 truecm}
\end{center}
\caption{(Color online) The same as in Fig.~\ref{fig:density_plus_up} except that $\phi_n=\pi$. The shapes vary from the inner circle to the external line as $k_\perp$ is increased from 0 to 2.0 GeV in steps of 0.25 GeV.}
\label{fig:density_minus_up}
\end{figure}

\begin{figure}[ht]
\begin{center}
\epsfig{file=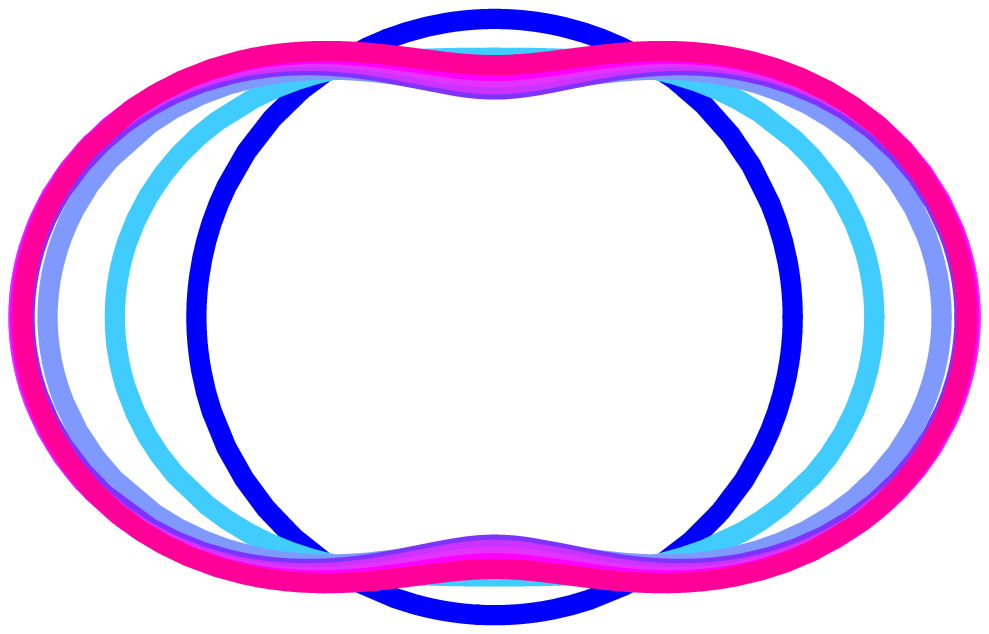,  width=7.5 truecm}
\hspace{1 truecm}
\epsfig{file=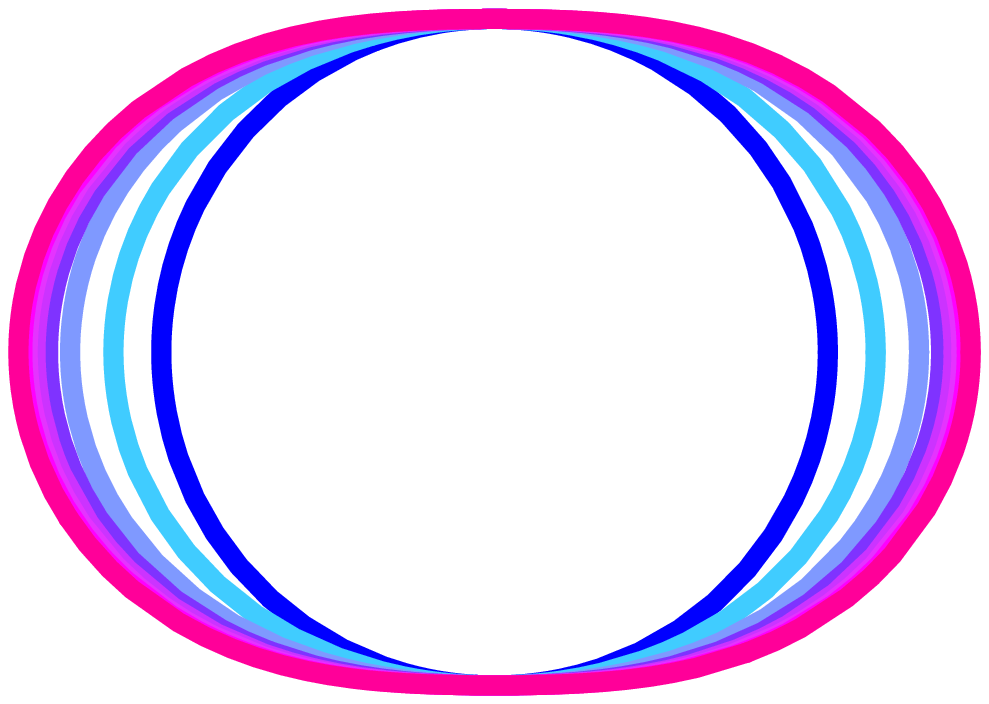,  width=7.5 truecm}
\end{center}
\caption{(Color online) Transverse shape of the proton,
$\hat \rho_{{\rm REL}\,T}(\boldsymbol{k}_\perp,\boldsymbol{n})/\tilde f_1(\boldsymbol{k}_\perp^2)$ 
assuming a struck $d$ quark. The horizontal axis is the direction of $\boldsymbol{S}_\perp$ and $\boldsymbol{n}=\hat{\boldsymbol{S}}_\perp$, $\phi_n=0$. The shapes vary from the inner circle to the external line as $k_\perp$ is increased from 0 to 2.0 GeV in steps of 0.25 GeV. The left figure is the results when only the contribution from the P waves to $\tilde h_{1T}^\perp$ is taken into account, and the right picture shows the total results when also the S and D waves are included. 
}
\label{fig:density_plus_down}
\end{figure}

\begin{figure}[ht]
\begin{center}
\epsfig{file=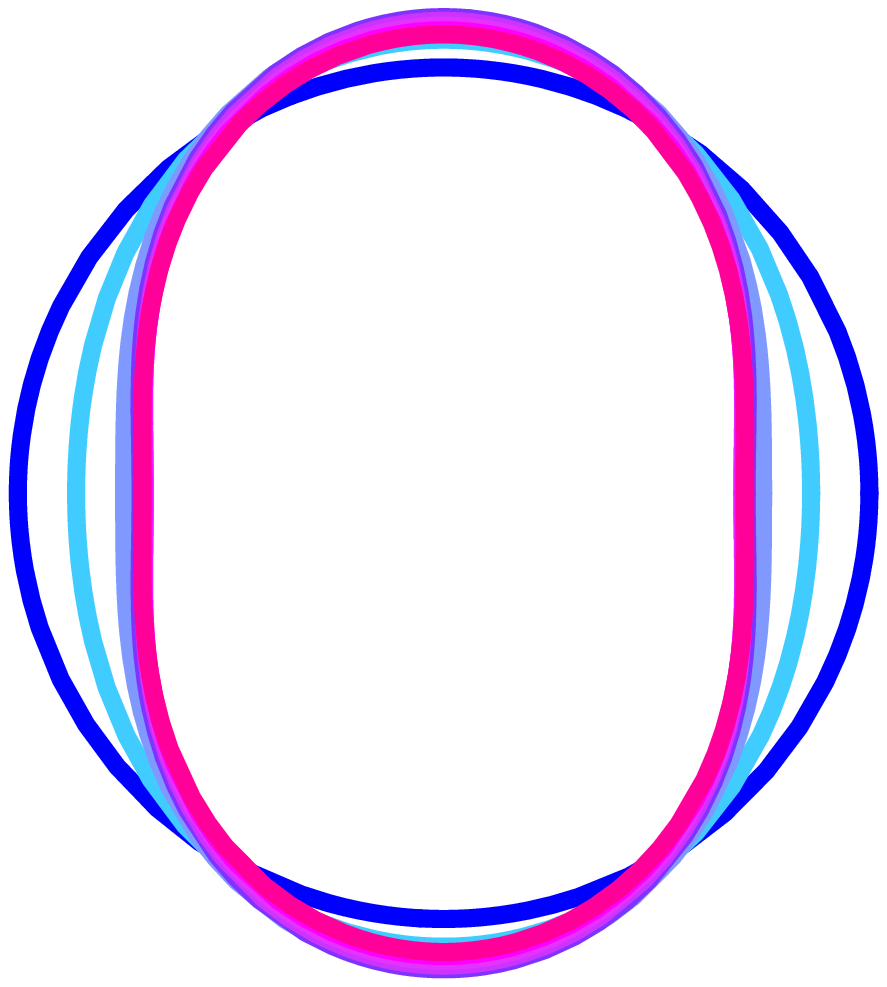,  width=7.5 truecm}
\hspace{1 truecm}
\epsfig{file=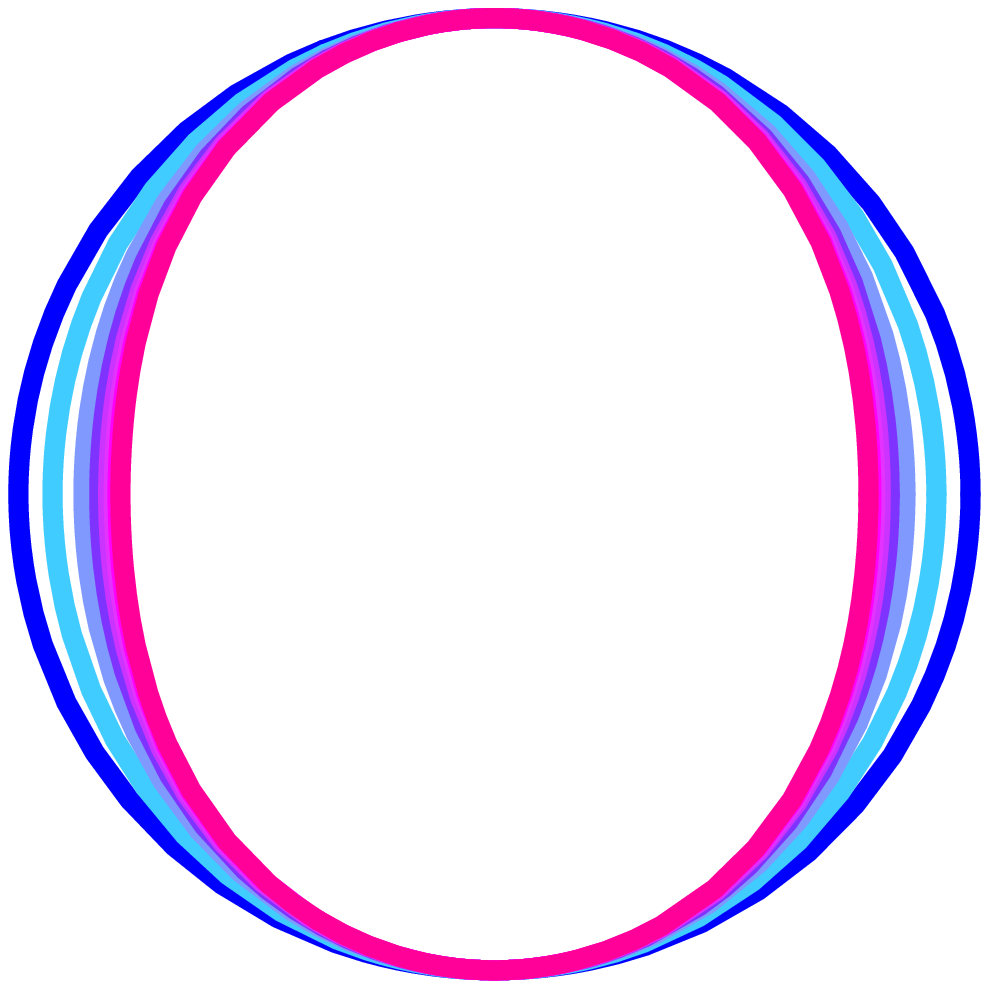,  width=7.5 truecm}
\end{center}
\caption{(Color online) the same as in Fig.~\ref{fig:density_plus_down} except that $\phi_n=\pi$. The shapes vary from the outer circle to the internal line as $k_\perp$ is increased from 0 to 2.0 GeV in steps of 0.25 GeV.}
\label{fig:density_minus_down}
\end{figure}

Assuming a struck $u$ quark, the transverse shapes of the proton are shown in Fig.~\ref{fig:density_plus_up} for $\boldsymbol{S}_T$ parallel to $\boldsymbol{n}$, $\phi_n=0$, and in Fig.~\ref{fig:density_minus_up} for $\boldsymbol{S}_T$ antiparallel to $\boldsymbol{n}$, $\phi_n=\pi$. The corresponding results assuming a struck $d$ quark are shown in Figs.~\ref{fig:density_plus_down} and \ref{fig:density_minus_down}, respectively. In our model $\tilde f_1^u$ ($\tilde f_1^d$) and $\tilde h_1^u$ (${\tilde h}_1^d$) are of the same (opposite) sign and similar size, so that the contribution of the first two terms on the rhs of Eq.~(\ref{eq:rhot}) tend to cancel each other for $\phi_n=\pi$ ($\phi_n=0$) emphasizing the role of the pretzelosity in producing deformation. For $u$ ($d$) quarks the last term in Eq.~(\ref{eq:rhot}) is  negative (positive) for $\phi=\phi_n=0$ and its size increases (reduces) with the inclusion of the D-wave. This explains the larger  transverse deformation in the direction antiparallel (parallel) to $\boldsymbol{S}_T$ for a struck $u$ ($d$) quark, with a more significant effect in the case of the $u$ quark.


\section{Concluding remarks}
\label{sect:conclude}

Quite general expressions have been derived for the transverse momentum dependent parton distributions in a light-cone description of the nucleon where the Fock expansion is truncated to consider only valence quarks. They are given in terms of matrix elements between nucleon states with different orbital angular momentum so that one can immediately appreciate the origin of each individual TMD. In particular, combining the total parton light-cone helicity with the nucleon helicity, angular momentum conservation requires the angular momentum projection of the nucleon to run from $-1$ to $2$, thus imposing S-, P- and D-wave components in the nucleon wave function. 

The complete three-quark light-cone wave function involves six amplitudes that can be either real or complex depending on the gauge fixing conditions~\cite{Ji:2002xn,Ji:2004}.  When complex, they incorporate the effects of final state interactions. The light-cone amplitudes have been used to obtain a model independent light-cone amplitude overlap representation of the T-even and T-odd TMDs, which emphasizes the role of the different angular momentum components. In the present approach where gluons are not included, the six amplitudes are real and  have been constructed by assuming a light-cone constituent quark model with SU(6) spin-flavor symmetry. Analytic formulae have been derived for the T-even TMDs in terms of the momentum dependent part of the LCWF. By simple inspection of such formulae we have found that among the six twist-two T-even TMDs there are three relations, so that only three TMDs are in fact independent when assuming SU(6) spin-flavor symmetry in the three-quark sector of the Fock expansion. Two of such relations, Eqs.~(\ref{eq:61}) and (\ref{eq:61a}), were already known~\cite{PPB07,Avakian08}, whereas the third one, Eq.~(\ref{eq:61b}), was first found in the diquark spectator model~\cite{Mulders3} and shown here to be valid in a larger class of relativistic quark models with SU(6) symmetry.

Numerical calculations of the T-even TMDs have been presented by adopting a power-law  form of the momentum dependent part of the LCWF. All distributions have either a maximum or minimum at $x\sim 0.2$  and $\boldsymbol{k}_\perp^2=0$ moving to higher values of $x$ with increasing $\boldsymbol{k}_\perp^2$. This indicates that the usual Gaussian ansatz for a factorized $\boldsymbol{k}_\perp^2$ dependence of TMDs is not adequate. In any case, the formalism described here is well suited to study the effects of the momentum dependence of the nucleon wave function, in particular in phenomenological applications to SIDIS and DY processes.

The role of the different orbital angular momentum components in the nucleon wave function is best appreciated by looking at the so-called pretzelosity, $h_{1T}^\perp$, and its effect on the spin-dependent quark density producing a nonspherical shape of the nucleon~\cite{Miller08}. A larger deformation in the direction antiparallel (parallel) to the transverse spin $\boldsymbol{S}_T$ is found for a struck $u$ ($d$) quark, with a significant sensitivity to the presence of a D-wave component in the nucleon wave function.

\bigskip

\begin{center}
\textbf{Acknowledgements}
\end{center}

The authors are grateful to Peter Schweitzer for providing them the numerical results of the pretzelosity in the bag model of Ref.~\cite{Avakian08} and for interesting discussions.


\begin{appendix}

\section{LCWF overlap representation of TMDs}

\label{sect:app}

Explicit expressions of the T-even TMDs in terms of overlap of the light-cone amplitudes of Section~\ref{sect:ampli} are given in this Appendix. The results for $g_{1T}^q$ and $h_{1L}^{\perp\,q}$ have already been derived in Ref.~\cite{Ji:2002xn}, while the results for the other T-even TMDs are given here for the first time. However, our expressions for  $g_{1T}^q$ and $h_{1L}^{\perp\,q}$ differ from the previous results of Ref.~\cite{Ji:2002xn} because we have corrected the wave function component with $\psi^{(5)}$ according to Ref.~\cite{Ji:2004}.

The TMD $f_1^q(x,\boldsymbol{ k}^2_\perp)$ reads
\begin{eqnarray}
\label{eq:50}
f_1^q(x,\boldsymbol{ k}^2_\perp)&=&\int d[1]d[2]d[3]\sqrt{x_1x_2x_3}\,
\,\mathcal{
F}^{q},
\end{eqnarray}
where for the up quark we have
\begin{eqnarray}
&&\mathcal{F}^{u}=
2\,\delta^3(k-k_1)
\tilde \psi^{(1,2)}(1,2,3)
[\tilde \psi^{(1,2)*}(1,2,3)
+\tilde \psi^{(1,2)*}(3,2,1)]\nn\\
&&+[\delta^3(k-k_3)+\delta^3(k-k_2)]
[\tilde \psi^{(1,2)}(1,2,3)\tilde \psi^{(1,2)*}(1,2,3)]\nn\\
& &+2\,\delta^3(k-k_2)\tilde \psi^{(3,4)}(1,2,3)
[\tilde \psi^{(3,4)*}(1,2,3)
+\tilde \psi^{(3,4)*}(1,3,2)]\nn\\
&&+[\delta^3(k-k_1)+\delta^3(k-k_3)]
[\tilde \psi^{(3,4)}(1,2,3)\tilde \psi^{(3,4)*}(1,2,3)]\nn\\
&&+[\delta(k-k_1)+\delta(k-k_2)]
[\tilde\psi^{(5)}(1,2,3)+\tilde\psi^{(5)}(2,1,3)]
[\tilde\psi^{(5)*}(1,2,3)-\tilde\psi^{(5)*}(1,3,2)]\nn\\
&&+[\delta(k-k_1)+\delta(k-k_3)]
[\tilde\psi^{(5)}(1,2,3)+\tilde\psi^{(5)}(3,2,1)]
[\tilde\psi^{(5)*}(1,2,3)-\tilde\psi^{(5)*}(1,3,2)]\nn\\
& &+[\delta^3(k-k_1)+\delta^3(k-k_3)]
[\tilde\psi^{(6)}(1,2,3)+\tilde\psi^{(6)}(3,2,1)]
[\tilde \psi^{(6)*}(1,2,3)-\tilde \psi^{(6)*}(1,3,2)],\nn\\
& &+[\delta^3(k-k_1)+\delta^3(k-k_2)]
[\tilde\psi^{(6)}(1,2,3)+\tilde\psi^{(6)}(2,1,3)]
[\tilde \psi^{(6)*}(1,2,3)-\tilde \psi^{(6)*}(1,3,2)],
\end{eqnarray}
with $\delta^3(k-k_i)=\delta(x-x_i)\delta(\boldsymbol{ k}_{\perp}-\boldsymbol{ k}_{i\perp})$ and
\begin{eqnarray}
& &\tilde\psi^{(1,2)}(1,2,3)=\psi^{(1)}(1,2,3)+i\epsilon_T^{\alpha\beta}k_{1\alpha}k_{2\beta}\psi^{(2)}(1,2,3),\label{eq:52a}\\
& &\tilde\psi^{(3,4)}(1,2,3)=k^+_1\psi^{(3)}(1,2,3)+k_{2}^+\psi^{(4)}(1,2,3),
\label{eq:52b}\\
& &\tilde\psi^{(5)}(1,2,3)=k^-_2\psi^{(5)}(1,2,3),\label{eq:52c}\\
&&\tilde \psi^{(6)}(1,2,3)=k^+_1k^+_3\psi^{(6)}(1,2,3).
\label{eq:52d}
\end{eqnarray}
For the down quark, we obtain
\begin{eqnarray}
&&\mathcal{F}^{d}=
\delta^3(k-k_2)
\tilde \psi^{(1,2)}(1,2,3)
[\tilde \psi^{(1,2)*}(1,2,3)
+\tilde \psi^{(1,2)*}(3,2,1)]\nn\\
&&+\delta^3(k-k_3)
[\tilde \psi^{(1,2)}(1,2,3)\tilde \psi^{(1,2)*}(1,2,3)]\nn\\
& &+
\delta^3(k-k_1)\tilde \psi^{(3,4)}(1,2,3)
[\tilde \psi^{(3,4)*}(1,2,3)
+\tilde \psi^{(3,4)*}(1,3,2)]\nn\\
&&+\delta^3(k-k_3)
[\tilde \psi^{(3,4)}(1,2,3)\tilde \psi^{(3,4)*}(1,2,3)]\nn\\
&&+\delta^3(k-k_3)
[\tilde
\psi^{(5)}(1,2,3)+\tilde\psi^{(5)}(2,1,3)]
[\tilde\psi^{(5)*}(1,2,3)-\tilde\psi^{(5)*}(1,3,2)]\nn\\
&&+\delta^3(k-k_2)
[\tilde
\psi^{(5)}(1,2,3)+\tilde\psi^{(5)}(3,2,1)]
[\tilde\psi^{(5)*}(1,2,3)-\tilde\psi^{(5)*}(1,3,2)]\nn\\
& &
+\delta^3(k-k_2)
[\tilde\psi^{(6)}(1,2,3)+\tilde\psi^{(6)}(3,2,1)]
[\tilde \psi^{(6)*}(1,2,3)-\tilde \psi^{(6)*}(1,3,2)],\nn\\
& &
+\delta^3(k-k_3)
[\tilde\psi^{(6)}(1,2,3)+\tilde\psi^{(6)}(2,1,3)]
[\tilde \psi^{(6)*}(1,2,3)-\tilde \psi^{(6)*}(1,3,2)].
\label{eq:53}
\end{eqnarray}

The result for $g_{1L}^q(x,\boldsymbol{ k}^2_\perp)$ reads
\begin{eqnarray}
\label{eq:54}
g_{1L}^q(x,\boldsymbol{ k}_\perp^2)&=&\int d[1]d[2]d[3]\sqrt{x_1x_2x_3}\, 
\,\mathcal{G}_L^{q},
\end{eqnarray}
where the function $\mathcal{G}_L^q$ for the up quark is 
\begin{eqnarray}
&&\mathcal{G}_L^{u}=
2\,\delta^3(k-k_1)
\tilde \psi^{(1,2)}(1,2,3)[\tilde \psi^{(1,2)*}(1,2,3)
+\tilde \psi^{(1,2)*}(3,2,1)]\nn\\
&&+[\delta^3(k-k_3)-\delta^3(k-k_2)]
[\tilde \psi^{(1,2)}(1,2,3)\tilde \psi^{(1,2)*}(1,2,3)]\nn\\
& &-
2\,\delta^3(k-k_2)\tilde \psi^{(3,4)}(1,2,3)
[\tilde \psi^{(3,4)*}(1,2,3)
+\tilde \psi^{(3,4)*}(1,3,2)]\nn\\
&&+[\delta^3(k-k_1)-\delta^3(k-k_3)]
[\tilde \psi^{(3,4)}(1,2,3)\tilde \psi^{(3,4)*}(1,2,3)]\nn\\
&&+[\delta(k-k_1)+\delta(k-k_2)]
[\tilde\psi^{(5)}(1,2,3)+\tilde\psi^{(5)}(2,1,3)]
[\tilde\psi^{(5)*}(1,2,3)-\tilde\psi^{(5)*}(1,3,2)]\nn\\
&&+[\delta(k-k_1)+\delta(k-k_3)]
[\tilde\psi^{(5)}(1,2,3)+\tilde\psi^{(5)}(3,2,1)]
[\tilde\psi^{(5)*}(1,2,3)-\tilde\psi^{(5)*}(1,3,2)]\nn\\
& &
-[\delta^3(k-k_1)+\delta^3(k-k_3)]
[\tilde\psi^{(6)}(1,2,3)+\tilde\psi^{(6)}(3,2,1)]
[\tilde \psi^{(6)*}(1,2,3)-\tilde \psi^{(6)*}(1,3,2)],\nn\\
& &
-[\delta^3(k-k_1)+\delta^3(k-k_2)]
[\tilde\psi^{(6)}(1,2,3)+\tilde\psi^{(6)}(2,1,3)]
[\tilde \psi^{(6)*}(1,2,3)-\tilde \psi^{(6)*}(1,3,2)],
\label{eq:55}
\end{eqnarray}
and for the down quark is
\begin{eqnarray}
&&\mathcal{G}_L^{d}=
-\delta^3(k-k_2)
\tilde \psi^{(1,2)}(1,2,3)
[\tilde \psi^{(1,2)*}(1,2,3)
+\tilde \psi^{(1,2)*}(3,2,1)]\nn\\
&&+\delta^3(k-k_3)
[\tilde \psi^{(1,2)}(1,2,3)\tilde \psi^{(1,2)*}(1,2,3)]\nn\\
& &+
\delta^3(k-k_1)\tilde \psi^{(3,4)}(1,2,3)
[\tilde \psi^{(3,4)*}(1,2,3)
+\tilde \psi^{(1,2)*}(1,3,2)]\nn\\
&&-\delta^3(k-k_3)
[\tilde \psi^{(3,4)}(1,2,3)\tilde \psi^{(3,4)*}(1,2,3)]\nn\\
&&+\delta^3(k-k_3)
[\tilde
\psi^{(5)}(1,2,3)+\tilde\psi^{(5)}(2,1,3)]
[\tilde\psi^{(5)*}(1,2,3)-\tilde\psi^{(5)*}(1,3,2)]\nn\\
&&+\delta^3(k-k_2)
[\tilde
\psi^{(5)}(1,2,3)+\tilde\psi^{(5)}(3,2,1)]
[\tilde\psi^{(5)*}(1,2,3)-\tilde\psi^{(5)*}(1,3,2)]\nn\\
& &
-\delta^3(k-k_2)
[\tilde\psi^{(6)}(1,2,3)+\tilde\psi^{(6)}(3,2,1)]
[\tilde \psi^{(6)*}(1,2,3)-\tilde \psi^{(6)*}(1,3,2)],\nn\\
& &
-\delta^3(k-k_3)
[\tilde\psi^{(6)}(1,2,3)+\tilde\psi^{(6)}(2,1,3)]
[\tilde \psi^{(6)*}(1,2,3)-\tilde \psi^{(6)*}(1,3,2)].
\label{eq:56}
\end{eqnarray}

For $h_1^q$ we have
\begin{eqnarray}
h_{1}^q(x,\boldsymbol{ k}^2_\perp)&=&\int d[1]d[2]d[3]\sqrt{x_1x_2x_3}\,
{\rm Re}\mathcal{H}^q,
\label{eq:57}
\end{eqnarray}
where for the up quark
\begin{eqnarray}
\mathcal{H}^u&=&
2\,\left\{
\delta^3(k-k_1)
[\tilde \psi^{(1,2)}(1,3,2)+\tilde \psi^{(1,2)}(2,3,1)]
\tilde \psi^{\overline{(1,2)}*}(1,2,3)\right.
\nn\\
&&-\delta(k-k_1)
[\tilde\psi^{\overline{(5)}}(1,2,3)+\tilde\psi^{\overline{(5)}}(2,1,3)]
\tilde\psi^{(3,4)*}(1,2,3)
\nn\\
& &\left.
+\delta^3(k-k_2)
[\tilde\psi^{\overline{(5)}}(1,3,2)+\tilde\psi^{\overline{(5)}}(2,3,1)]
\tilde \psi^{(3,4)*}(2,1,3)\right\},
\label{eq:58}
\end{eqnarray}
with
\begin{eqnarray}
& &\tilde\psi^{\overline{(1,2)}}(1,2,3)=
\psi^{(1)}(1,2,3)-i\epsilon_T^{\alpha\beta}k_{1\alpha}k_{2\beta}\psi^{(2)}(1,2,3),\label{eq:59a}\\
& &\tilde\psi^{\overline{(5)}}(1,2,3)=k^+_2\psi^{(5)}(1,2,3).
\label{eq:59b}
\end{eqnarray}
For the down quark, we obtain
\begin{eqnarray}
\mathcal{H}^d&=&\left\{
-\delta^3(k-k_3)\tilde \psi^{(1,2)}(1,2,3)
\tilde \psi^{\overline{(1,2)}*}(2,1,3)\right.
\nn\\
&&+2\delta(k-k_3)
[\tilde\psi^{\overline{(5)}}(1,2,3)+\tilde\psi^{\overline{(5)}}(2,1,3)]
\tilde\psi^{(3,4)*}(3,2,1)
\nn\\
& &\left.
-2\delta^3(k-k_2)
[\tilde\psi^{\overline{(5)}}(1,2,3)+\tilde\psi^{\overline{(5)}}(3,2,1)]
\tilde \psi^{(3,4)*}(2,1,3)\right\}\nn\\
&&=\delta^3(k-k_3)\,{\rm Re}\left\{
-\tilde \psi^{(1,2)}(1,2,3)
\tilde \psi^{\overline{(1,2)}*}(2,1,3)
\right.\nn\\
&&+2
[\tilde\psi^{\overline{(5)}}(1,2,3)+\tilde\psi^{\overline{(5)}}(2,1,3)]
\tilde\psi^{(3,4)*}(3,2,1)
\nn\\
& &\left.
-2
[\tilde\psi^{\overline{(5)}}(1,3,2)+\tilde\psi^{\overline{(5)}}(2,3,1)]
\tilde \psi^{(3,4)*}(3,1,2)\right\}.
\label{eq:60}
\end{eqnarray}

The result for $g_{1T}^q$ is
\begin{eqnarray}
\label{eq:62}
&&g_{1T}^q(x,\boldsymbol{ k}^2_{\perp})=
\frac{M}{\boldsymbol{k}_{\perp}^{2}}
\int d[1]d[2]d[3]\sqrt{x_1x_2x_3}\,
{\rm Re}\mathcal{G}_{T}^{q}.
\end{eqnarray}
The function $\mathcal{G}_T^q$ for up quarks is
\begin{eqnarray}
\mathcal{G}_{T}^{u}&=&
2\left\{\delta^3(k-k_1)   \psi^{(1,2)*}_j(1,2,3)\psi^{(3,4)}_j(2,1,3)
\right.
\nn\\
&&\qquad
-\delta^3(k-k_2)    \psi^{(1,2)}_j(1,2,3) \psi^{(3,4)*}_j(2,1,3)\nn\\
&&\qquad
+[\delta^3(k-k_1)+\delta^3(k-k_3)]   \psi^{(1,2)*}_j(1,2,3)
[ \psi^{(3,4)}_j(2,1,3)+ \psi^{(3,4)}_j(2,3,1)]\nn\\
&&\qquad 
\left.+[\delta^3(k-k_1)+\delta^3(k-k_2)][\psi^{(5)*}(1,2,3)\psi^{(6+)}(2,1,3)\right.\nn\\
&&\hspace{3cm}\left.
+\psi^{(5)*}(1,3,2)\psi^{(6+')}(3,2,1) ]\right\},
\label{eq:63}
\end{eqnarray}
where
\begin{eqnarray}
\label{eq:64}
  \psi_{j}^{(1,2)}(1,2,3)&=&
\psi^{(1)}(1,2,3)k^{j}-\epsilon_T^{ij}k^{i}(k_{1}^{x}k_{2}^{y}-k_{1}^{y}k_{2}^{x})
\psi^{(2)}(1,2,3),
\\
\label{eq:65}
\psi_{j}^{(3,4)}(1,2,3)&=&k_{1}^{j}
\psi^{(3)}(1,2,3)+k_{2}^{j}\psi^{(4)}(1,2,3),\\
\psi^{(6+')}(1,2,3)&=&
\left(\boldsymbol{k}_{2\perp}^{2}\boldsymbol{k}_{3\perp}\cdot \boldsymbol{k}_{\perp}
+ \boldsymbol{k}_{3\perp}^{2}\boldsymbol{k}_{2\perp}\cdot \boldsymbol{k}_{\perp}\right)
\left(\psi^{(6)}(2,1,3)+\psi^{(6)}(3,1,2)\right)
\nonumber\\
&&+ \boldsymbol{k}_{1\perp}^{2}\boldsymbol{k}_{2\perp}\cdot \boldsymbol{k}_{\perp}
\psi^{(6)}(2,3,1)+ \boldsymbol{k}_{1\perp}^{2}\boldsymbol{k}_{3\perp}
\cdot \boldsymbol{k}_{\perp}\psi^{(6)}(3,2,1),
\label{eq:66}\\
\psi^{(6+)}(1,2,3)&=&
-\boldsymbol{k}_{1\perp}^{2}\boldsymbol{k}_{3\perp}\cdot \boldsymbol{k}_{\perp}
\psi^{(6)}(1,2,3)\nonumber\\
&&+\left(\boldsymbol{k}_{2\perp}^{2}\boldsymbol{k}_{3\perp}\cdot 
\boldsymbol{k}_{\perp}+ \boldsymbol{k}_{3\perp}^{2}\boldsymbol{k}_{2\perp}\cdot \boldsymbol{k}_{\perp}\right)\psi^{(6)}(2,1,3)\nonumber\\
&&+ \boldsymbol{k}_{1\perp}^{2}\boldsymbol{k}_{2\perp}\cdot \boldsymbol{k}_{\perp}\left(\psi^{(6)}(1,3,2)+\psi^{(6)}(2,3,1)\right),
\label{eq:67}
\end{eqnarray}
with $i,j=x,y$.
For the down quark we obtain
\begin{eqnarray}
\mathcal{G}_{T}^{d}&=&
2\left\{\delta^3(k-k_3)    \psi^{(1,2)*}_j(1,2,3) \psi^{(3,4)}_j(2,1,3)\right.\nn\\
& &\quad{}
-\delta^3(k-k_2)    \psi^{(1,2)}_j(1,2,3)
[ \psi^{(3,4)*}_j(2,1,3)+ \psi^{(3,4)*}_j(2,3,1)]
\nn\\
& &\quad{}
\left.+\delta^3(k-k_3)
[\psi^{(5)*}(1,2,3)\psi^{(6+)}(2,1,3) +\psi^{(5)*}(1,3,2)\psi^{(6+')}(3,2,1)
]\right\}.\label{eq:68}
\end{eqnarray}
As outlined in Ref.~\cite{Ji:2002xn} the imaginary part of the function $\mathcal{G}_T^q$ in Eq.~(\ref{eq:62}) gives  the corresponding LCWF overlap representation for the T-odd Sivers TMD $f_{1T}^{\perp\,q}$~\cite{Sivers}.

The result for $h_{1L}^{\perp\,q}$ is
\begin{eqnarray}
&&
h_{1L}^{\perp\,q}(x,\boldsymbol{k}^2_{\perp})
=\frac{M}{\Lambda\boldsymbol{k}_{\perp}^{2}}\int d[1]d[2]d[3]\sqrt{x_1x_2x_3}
\,{\rm Re}\,
\mathcal{H}_{L}^{\perp \,q},\label{eq:69}
\end{eqnarray}
where the function $H_L^{\perp \,q}$ for the up quark is
\begin{eqnarray}
&&\mathcal{H}_{L}^{\perp \,u}=2\left\{
-\delta^3(k-k_1)
   \psi^{(1,2)'}_j(1,2,3)
[  \psi^{(3,4)*}_j(3,2,1)
+  \psi^{(3,4)*}_j(3,1,2)]\right.\nn\\
&&
-\delta^3(k-k_2)
[    \psi^{(1,2)'}_j(1,3,2)
+    \psi^{(1,2)'}_j(2,3,1)]
  \psi^{(3,4)*}_j(1,2,3)
\nn\\
&&
-\delta^3(k-k_2)
[  \psi^{(5-)}_j(1,2,3)
+  \psi^{(5-)}_j(2,1,3)]
  \psi^{(1,2)*}_j(1,2,3)
\nn\\
&&+\delta^3(k-k_1)
[  \psi^{(6,+)*}(1,2,3)\psi^{(3)}(1,2,3)
+  \psi^{(6,+)*}(2,1,3)\psi^{(4)}(1,2,3)],
\label{eq:70}
\end{eqnarray}
where 
\begin{eqnarray}
  \psi^{(5-)}_j(1,2,3)&=&
k_{2}^j\psi^{(5)}(1,2,3)
-k_{3}^j\psi^{(5)}(1,3,2),
\label{eq:71}\\
\psi^{(1,2)'}_j(1,2,3)&=&k^j\psi^{(1)}(1,2,3)+\epsilon_T^{ij}k^i
(k_1^xk_2^y-k_1^yk_2^x)\psi^{(2)}(1,2,3).
\label{eq:72}
\end{eqnarray}
The result for the down quark is
\begin{eqnarray}
&&\mathcal{H}_{L}^{\perp\,d}=
2\left\{
\delta^3(k-k_3)
    \psi^{(1,2)'}_j(1,2,3) \psi^{(3,4)*}_j(1,2,3)\right.
\nn\\
&&-\delta^3(k-k_2)
    \psi^{(1,2)*}_j(1,2,3)
[  \psi^{(5-)}_j(1,2,3)
+  \psi^{(5-)}_j(3,2,1)]
\nn\\
&&\left.+\delta^3(k-k_1)
[  \psi^{(6,+)'*}(1,2,3)\psi^{(3)}(1,2,3)
-  \psi^{(6,+)*}(2,3,1)\psi^{(4)}(1,2,3)]\right\}.
\label{eq:73}
\end{eqnarray}
Note that by taking the imaginary part of the function $\mathcal{H}^{\perp\, q}_L$ in Eq.~(\ref{eq:69}) one obtains the corresponding results for the T-odd Boer-Mulders function $h_{1}^{\perp\,q}$~\cite{BoerMulders}.

Finally, for the $h_{1T}^{\bot\,q}$ TMD we obtain
\begin{eqnarray}
h_{1T}^{\bot\,q}(x,\boldsymbol{k}^2_{\perp})=\frac{2M^2}{k^2_y-k^2_x}\int d[1]d[2]d[3]\sqrt{x_1x_2x_3}
\,{\rm Re}\,
\mathcal{H}_{T}^{\bot\,q},\label{eq:74}
\end{eqnarray}
where $\mathcal{H}_{T}^{\bot\,q}$ is given by
\begin{eqnarray}
\mathcal{H}_{T}^{\bot\,u}&=&
2\,\left\{
\delta(k-k_2)
\tilde\psi^{(3,4)}(1,2,3)
[\tilde\psi^{\overline{(3,4)}*}(3,2,1)
+\tilde\psi^{\overline{(3,4)}*}(3,1,2)
]
\right.
\nn\\
&&
+\delta^3(k-k_1)
[\tilde\psi^{(6)}(1,2,3)+\tilde\psi^{(6)}(3,2,1)]
\tilde \psi^{\overline{(1,2)}*}(3,1,2)
\nn\\
&&\left.
-\delta^3(k-k_2)
[\tilde\psi^{(6)}(1,2,3)+\tilde\psi^{(6)}(2,1,3)]
\tilde \psi^{\overline{(1,2)}*}(1,2,3)
\right\},
\label{eq:75}
\end{eqnarray}
with
\begin{eqnarray}
& &\tilde\psi^{\overline{(3,4)}}(1,2,3)=k^-_1\psi^{(3)}(1,2,3)+k_{2}^-\psi^{(4)}(1,2,3).\label{eq:77}
\end{eqnarray}
For the down quark, one finds:
\begin{eqnarray}
\mathcal{H}_{T}^{\bot\,d}&=&
-\left\{\delta^3(k-k_3)
[\tilde\psi^{(3,4)}(1,2,3)
\tilde\psi^{\overline{(3,4)}*}(2,1,3)]\right.
\nn\\
&&
+2\delta^3(k-k_2)
[\tilde\psi^{(6)}(1,2,3)-\tilde\psi^{(6)}(3,1,2)]
\tilde \psi^{\overline{(1,2)}*}(1,2,3)
\nn\\
&&\left.
+2\delta^3(k-k_3)
[\tilde\psi^{(6)}(2,3,1)-\tilde\psi^{(6)}(1,2,3)]
\tilde \psi^{\overline{(1,2)}*}(1,3,2)
\right\}.
\label{eq:78}
\end{eqnarray}

\end{appendix}


\clearpage


\clearpage




\begin{thebibliography}{35}

\expandafter\ifx\csname natexlab\endcsname\relax\def\natexlab#1{#1}\fi
\expandafter\ifx\csname bibnamefont\endcsname\relax
  \def\bibnamefont#1{#1}\fi
\expandafter\ifx\csname bibfnamefont\endcsname\relax
  \def\bibfnamefont#1{#1}\fi
\expandafter\ifx\csname citenamefont\endcsname\relax
  \def\citenamefont#1{#1}\fi
\expandafter\ifx\csname url\endcsname\relax
  \def\url#1{\texttt{#1}}\fi
\expandafter\ifx\csname urlprefix\endcsname\relax\def\urlprefix{URL }\fi
\providecommand{\bibinfo}[2]{#2}
\providecommand{\eprint}[2][]{\url{#2}}

\bibitem{Collins:1981uk}
J.~C.~Collins and D.~E. Soper, Nucl.\ Phys.\  {\bf B 193}, 381 (1981); Erratum-\emph{ibid.}\  \textbf{B 213}, 545 (1983)].

\bibitem{Ji:2004wu}
X.~Ji, J.-P.~Ma and F.~Yuan, Phys.\ Rev.\  \textbf{D 71}, 034005 (2005); Phys.\ Lett.\ \textbf{B 597}, 299 (2004).

\bibitem{Collins:2004nx}
J.~C.~Collins and A.~Metz, Phys.\ Rev.\ Lett.\  \textbf{93}, 252001 (2004).

\bibitem{Collins:2003fm}
J.~C.~Collins, Acta Phys.\ Polon.\  \textbf{B 34}, 3103 (2003).

\bibitem{Collins:2007ph}
J.~C.~Collins, T.~C.~Rogers and A.~M.~Stasto, Phys.\ Rev.  \textbf{D 77}, 085009 (2008).

\bibitem{Bom08}
C.~J.~Bomhof, P.J.~Mulders, Nucl. Phys. \textbf{B 795}, 409 (2008).

\bibitem{Mulders2} 
P.J. Mulders, R.D. Tangerman, Nucl. Phys. \textbf{B 461}, 197 (1996); Erratum-\emph{ibid.} \textbf{B 484}, 538 (1997).

\bibitem{BoerMulders} 
D. Boer, P.J. Mulders, Phys. Rev. \textbf{D 57}, 5780 (1998).

\bibitem{Goeke} 
K. Goeke, A. Metz, M. Schlegel, Phys. Lett. \textbf{B 618}, 90 (2005).

\bibitem{Bacchetta} 
A.~Bacchetta, M.~Diehl, K.~Goeke, A.~Metz, P.~J.~Mulders and M.~Schlegel, JHEP \textbf{02}, 093  (2007).

\bibitem{LB80}
G.P.~Lepage, S.J.~Brodsky, Phys. Rev. \textbf{D 22}, 2157 (1980).

\bibitem{BPP98}
S.J.~Brodsky, H.-Ch.~Pauli, S.S.~Pinsky, Phys. Rep. \textbf{301}, 299 (1998).

\bibitem[{\citenamefont{Chernyak and
  Zhitnitsky}(1984{\natexlab{a}})}]{Chernyak:1984ej}
\bibinfo{author}{\bibfnamefont{V.~L.} \bibnamefont{Chernyak}} \bibnamefont{and}
  \bibinfo{author}{\bibfnamefont{A.~R.} \bibnamefont{Zhitnitsky}},
  \bibinfo{journal}{Phys. Rep.} \textbf{\bibinfo{volume}{112}},
  \bibinfo{pages}{173} (\bibinfo{year}{1984}{\natexlab{a}}).

\bibitem[{\citenamefont{King and Sachrajda}(1987)}]{King:1987wi}
\bibinfo{author}{\bibfnamefont{I.~D.} \bibnamefont{King}} \bibnamefont{and}
  \bibinfo{author}{\bibfnamefont{C.~T.} \bibnamefont{Sachrajda}},
  \bibinfo{journal}{Nucl. Phys.} \textbf{\bibinfo{volume}{B 279}},
  \bibinfo{pages}{785} (\bibinfo{year}{1987}).

\bibitem[{\citenamefont{Chernyak et~al.}(1989)\citenamefont{Chernyak, Ogloblin,
  and Zhitnitsky}}]{Chernyak:1989nv}
\bibinfo{author}{\bibfnamefont{V.~L.} \bibnamefont{Chernyak}},
  \bibinfo{author}{\bibfnamefont{A.~A.} \bibnamefont{Ogloblin}},
  \bibnamefont{and} \bibinfo{author}{\bibfnamefont{I.~R.}
  \bibnamefont{Zhitnitsky}}, \bibinfo{journal}{Z. Phys.}
  \textbf{\bibinfo{volume}{C 42}}, \bibinfo{pages}{583} (\bibinfo{year}{1989}).

\bibitem[{\citenamefont{Braun et~al.}(1999)\citenamefont{Braun, Derkachov,
  Korchemsky, and Manashov}}]{Braun:1999te}
\bibinfo{author}{\bibfnamefont{V.~M.} \bibnamefont{Braun}},
  \bibinfo{author}{\bibfnamefont{S.~E.} \bibnamefont{Derkachov}},
  \bibinfo{author}{\bibfnamefont{G.~P.} \bibnamefont{Korchemsky}},
  \bibnamefont{and} \bibinfo{author}{\bibfnamefont{A.~N.}
  \bibnamefont{Manashov}}, \bibinfo{journal}{Nucl. Phys.}
  \textbf{\bibinfo{volume}{B 553}}, \bibinfo{pages}{355} (\bibinfo{year}{1999}).
  \eprint[http://arXiv.org/abs]

\bibitem[{\citenamefont{Braun et~al.}(2000)\citenamefont{Braun, Fries, Mahnke,
  and Stein}}]{Braun:2000kw}
\bibinfo{author}{\bibfnamefont{V.}~\bibnamefont{Braun}},
  \bibinfo{author}{\bibfnamefont{R.~J.} \bibnamefont{Fries}},
  \bibinfo{author}{\bibfnamefont{N.}~\bibnamefont{Mahnke}}, \bibnamefont{and}
  \bibinfo{author}{\bibfnamefont{E.}~\bibnamefont{Stein}},
  \bibinfo{journal}{Nucl. Phys.} \textbf{\bibinfo{volume}{B 589}},
  \bibinfo{pages}{381} (\bibinfo{year}{2000}); Erratum-\emph{ibid.}\  \textbf{B 607}, 433 (2001)
  \eprint[http://arXiv.org/abs]
  
\bibitem{Stefanis}
N.G. Stefanis, Eur. Phys.  J. direct \textbf{C 7}, 1 (1999).

\bibitem{BurkJiY02}
M.~Burkardt, X.~Ji, F.~Yuan, Phys. Lett. \textbf{B 545}, 345 (2002).

\bibitem[{\citenamefont{Ji et~al.}(2003{\natexlab{a}})\citenamefont{Ji, Ma, and
  Yuan}}]{Ji:2002xn}
\bibinfo{author}{\bibfnamefont{X.}~\bibnamefont{Ji}},
  \bibinfo{author}{\bibfnamefont{J.-P.} \bibnamefont{Ma}}, \bibnamefont{and}
  \bibinfo{author}{\bibfnamefont{F.}~\bibnamefont{Yuan}},
  \bibinfo{journal}{Nucl. Phys.} \textbf{\bibinfo{volume}{B 652}},
  \bibinfo{pages}{383} (\bibinfo{year}{2003}{\natexlab{a}}).

\bibitem[{\citenamefont{Ji et~al.}(2004{\natexlab{a}})\citenamefont{Ji, Ma, and
  Yuan}}]{Ji:2004}
\bibinfo{author}{\bibfnamefont{X.}~\bibnamefont{Ji}},
  \bibinfo{author}{\bibfnamefont{J.-P.} \bibnamefont{Ma}}, \bibnamefont{and}
  \bibinfo{author}{\bibfnamefont{F.}~\bibnamefont{Yuan}},
  \bibinfo{journal}{Eur. Phys. J.} \textbf{\bibinfo{volume}{C 33}},
  \bibinfo{pages}{75} (\bibinfo{year}{2004}{\natexlab{a}}).

\bibitem[{\citenamefont{Ji et~al.}(2003{\natexlab{a}})\citenamefont{Ji, Ma, and
  Yuan}}]{Ji:2003}
\bibinfo{author}{\bibfnamefont{X.}~\bibnamefont{Ji}},
  \bibinfo{author}{\bibfnamefont{J.-P.} \bibnamefont{Ma}}, \bibnamefont{and}
  \bibinfo{author}{\bibfnamefont{F.}~\bibnamefont{Yuan}},
  \bibinfo{journal}{Phys. Rev. Lett.} \textbf{\bibinfo{volume}{90}},
  \bibinfo{pages}{241601} (\bibinfo{year}{2003}{\natexlab{a}}).
  
\bibitem{PBff}
B. Pasquini, and S. Boffi, Phys. Rev. \textbf{D 76}, 074011 (2007).

\bibitem{BPT1}
S. Boffi, B. Pasquini and M. Traini, Nucl. Phys.  \textbf{B 649}, 243 (2003).

\bibitem{BPT2}
S. Boffi, B. Pasquini and M. Traini, Nucl. Phys.  \textbf{B 680}, 147 (2004)

\bibitem{PPB}
B. Pasquini, M. Pincetti, and S. Boffi, Phys. Rev.  \textbf{D 72}, 094029 (2005)

\bibitem{PBcloud}
B. Pasquini, and S. Boffi, Phys. Rev. \textbf{D 73}, 094001 (2006).

\bibitem{PBspin}
B. Pasquini, and S. Boffi, Phys. Lett.  \textbf{B 653}, 23 (2007)

\bibitem{Bom04}
C.J.~Bomhof, P.J.~Mulders, F.~Pijlman, Phys. Lett. \textbf {B 596}, 277 (2004); Eur. Phys. J. \textbf{C 47}, 147 (2006).

\bibitem{Sivers}
D.W.~Sivers, Phys. Rev. \textbf{D 41}, 83 (1990).

\bibitem{JiYuan:02}
X.~Ji, F.~Yuan, Phys. Lett. \textbf{B 543}, 66 (2002).

\bibitem{JiYuanBel:03}
A.V.~Belitsky, X.~Ji, F.~Yuan, Nucl. Phys. \textbf{B 656}, 165 (2003).

\bibitem{Mulders}
D. Boer, P.J.~Mulders, F.~Pijlman, Nucl. Phys. \textbf{B 667}, 201 (2007).

\bibitem[{\citenamefont{Franklin}(2004{\natexlab{a}})\citenamefont{Franklin}}]{Franklin:68}
\bibinfo{author}{\bibfnamefont{J.}~\bibnamefont{Franklin}},
  \bibinfo{journal}{Phys. Rev. } \textbf{\bibinfo{volume}{172}},
  \bibinfo{pages}{1807} (\bibinfo{year}{1968}{\natexlab{a}}).

\bibitem[{\citenamefont{Capstick}(1986{\natexlab{a}})\citenamefont{Capstick and Isgur}}]{Capstick:86}
\bibinfo{author}{\bibfnamefont{S.}~\bibnamefont{Capstick}}, \bibnamefont{and}
  \bibinfo{author}{\bibfnamefont{N.}~\bibnamefont{Isgur}},
  \bibinfo{journal}{Phys. Rev. } \textbf{\bibinfo{volume}{D 34}},
  \bibinfo{pages}{2809} (\bibinfo{year}{1986}{\natexlab{a}}).
  
\bibitem{CS08}
I.~O.~Cherednikov and N.~G.~Stefanis, Phys. Rev. \textbf{D 77}, 094001 (2008); arXiv:0802.2821 [hep-ph].

\bibitem{Melosh:74}
H.J.~Melosh, Phys. Rev.  \textbf{D 9}, 1095 (1974).

\bibitem{PPB07}
B. Pasquini, M. Pincetti, and S. Boffi, Phys. Rev.  \textbf{D 76}, 034020 (2007)

\bibitem{Avakian08}
H.~Avakian, A.~V.~Efremov, P.~Schweitzer, F.~Yuan, arXiv:0805.3355 [hep-ph].

\bibitem{Meissner} 
S. Meissner, A. Metz, and K. Goeke, Phys. Rev. \textbf{D 76}, 034002 (2007).

\bibitem{BPreview}
S.~Boffi and B.~Pasquini, Rivista Nuovo Cim. \textbf{30}, 387 (2007).

\bibitem{Schlumpf:94a}
F.~Schlumpf, doctoral thesis, University of Zurich, 1992; hep-ph/9211255.

\bibitem{Schlumpf:94b}
F.~Schlumpf, J. Phys. G: Nucl. Part. Phys. \textbf{20}, 237 (1994);
Phys. Rev. \textbf{D 47}, 4114 (1993); Erratum-\emph{ibid.}\  \textbf{D 49}, 6246 (1993);
S.J. Brodsky and F.~Schlumpf, Phys. Lett. \textbf{B 329}, 111 (1994).

\bibitem{Burkardt07a}
M.~Burkardt, arXiv:0709.2966 [hep-ph].

\bibitem{Avakian08a}
H.~Avakian, A.~V.~Efremov, K.~Goeke, A.~Metz, P.~Schweitzer, T.~Teckentrup, Phys. Rev. \textbf{D 77}, 014023 (2008).

\bibitem{Faccioli}
P. Faccioli, M. Traini, and V. Vento, Nucl. Phys. \textbf{A 656}, 400 (1999).

\bibitem{Giannini} 
M. Ferraris, M.M. Giannini, M. Pizzo, E. Santopinto, L. Tiator, Phys. Lett. \textbf{B 364}, 231 (1995).

\bibitem{Mulders3}
R. Jakob, P.J. Mulders, and J. Rodrigues, Nucl. Phys. \textbf{A 626}, 937 (1997).
\bibitem{BBHM}
A.~Bacchetta, M. Boglione, A.~Henneman, and P.J.~Mulders, Phys. Rev. Lett. \textbf{85}, 712 (2000).

\bibitem{Conti}
A. Bacchetta, F. Conti, and M. Radici,
e-Print: arXiv:0807.0323 [hep-ph].

\bibitem{Gamberg}
L.P. Gamberg, G.R. Goldstein, and M. Schlegel, Phys. Rev. D \textbf{ 77}, 094016
(2008).

 \bibitem{Kotzinian:2006dw}
    A.~Kotzinian, B.~Parsamyan and A.~Prokudin,
    Phys.\ Rev.\  D {\bf 73}, 114017 (2006).

\bibitem{Miller08}
G.A.~Miller, Phys. Rev. \textbf{C 76}, 065209 (2007).


\end{thebibliography}
\end{document}